\documentclass[1,12pt]{elsarticle}
\usepackage{graphicx}
\usepackage{graphicx,epstopdf}
\usepackage{caption}
\usepackage{subcaption}
\usepackage{float}
\usepackage{amssymb,amsmath}
\usepackage{exscale}
\usepackage{makeidx,shortvrb,latexsym}
\usepackage{epstopdf}
\usepackage{tabularx, booktabs, multirow}
\usepackage{array}
\usepackage[]{hyperref}
\usepackage{color}

\journal{Computational Mechanics}
\newcounter{defcounter}
\usepackage[bbgreekl]{mathbbol}

\begin{document}
\begin{frontmatter}
\title{On the accuracy of spectral solvers for micromechanics based fatigue modeling}
\author{S. Lucarini$^{1, 2 }$\corref{cor1}}
\author{J. Segurado$^{1, 2 }$\corref{cor2}}
\address{$^1$ IMDEA Materials Institute \\ C/ Eric Kandel 2, 28906, Getafe, Madrid, Spain. \\\ \\
$^2$ Department of Materials Science, Technical University of Madrid/Universidad Polit\'ecnica de Madrid \\ E. T. S. de Ingenieros de Caminos. 28040 - Madrid, Spain. }
\cortext[cor2]{Corresponding author at: IMDEA Materials Institute, Spain \\ E-mail address: javier.segurado@imdea.org (J. Segurado)} 
\begin{abstract}
A framework based on FFT is proposed for micromechanical fatigue modeling of polycrystals as alternative to the Finite Element method (FEM). The variational FFT approach \cite{Geers2016,Geers2017}  is used with a crystal plasticity model for the cyclic behavior of the grains introduced through a FEM material subroutine, in particular an \emph{Abaqus umat}. The framework also includes an alternative projection operator based on discrete differentiation to improve the microfield fidelity allowing to include second phases.

The accuracy and efficiency of the FFT framework for microstructure sensitive fatigue prediction are assessed by comparing with FEM. The macroscopic cyclic response of a polycrystal obtained with both methods were indistinguishable, irrespective of the number of cycles. The microscopic fields presented small differences that decrease when using the discrete projection operator, which indeed allowed simulating accurately microstructures containing very stiff particles. Finally, the maximum differences in the fatigue life estimation from the microfields respect FEM were around 15\% . In summary, this framework allows predicting fatigue life with a similar accuracy than using FEM but strongly reducing the computational cost.
\end{abstract}
\end{frontmatter}

\section*{Notation}
\begin{tabular}{ll}
$\Omega_0$ & Domain in reference configuration \\
$\mathbf{x},\boldsymbol{\xi},\boldsymbol{\alpha}$  & Vectors $x_{i},\xi_{i},\alpha_{i}$\\
$\mathbf{P},\mathbf{F},\boldsymbol{\tau}$ & Second-order tensors $P_{ij},F_{ij},\tau_{ij}$\\
$\mathbb{C},\mathbb{G},\mathbb{K}$ & Fourth-order tensors and operators $C_{ijkl},G_{ijkl},K_{ijkl}$\\
$\mathbf{A}=\mathbf{F}^T$ & Tensor transpose $A_{ij}=F_{ji}$ \\
$\mathbf{A}=\boldsymbol{\tau}\mathbf{F}$ & Dot product $A_{ij}=\tau_{ip}F_{pj}$\\
$a=\mathbf{F}:\mathbf{P}$ & Double dot product $a=F_{ij}P_{ij}$ \\
$\mathbf{P}=\mathbb{C}:\mathbf{F}$ & Double dot product $P_{ij}=C_{ijkl}F_{kl}$ \\
$\mathrm{Div} \ \mathbf{P}$ & Divergence of tensor field in the reference configuration $\frac{\partial P_{ij}}{\partial X_j}$ \\
$G \ast P$ & Convolution operation \\
$I_{ijkl}$ & Fourth order identity tensor $I_{ijkl}=\delta_{ik}\delta_{jl}$ \\
\end{tabular}

\section{Introduction}
Micromechanics based fatigue models are essential tools for understanding the incubation and early propagation stages of fatigue cracks in engineering alloys \citep{MCDOWELL20101521,PINEAU2016484}. These models are based on computational polycrystalline homogenization techniques \citep{SLLL18, HerreraSolaz2014,Cruzado2015} in which the macroscopic response and the microscopic fields of a polycrystal subjected to a given load history are obtained by solving a boundary value problem on a representative volume element (RVE) of the microstructure. The RVE of the polycrystalline metal consists in an aggregate of grains with size, shape and orientation distributions representative of the microstructure considered. The mechanical behavior of each grain is modeled using a crystal plasticity model that includes the relevant features of cyclic plasticity such as kinematic hardening, ratcheting or cyclic softening \citep{CAILLETAUD199255,Cruzado2017}.  Within this framework, the heterogeneous distribution of stress, strains and other internal variable microfields are resolved, allowing the use of local fatigue criteria to estimate the fatigue life of the alloy  \citep{Shenoy2007,McDowell2010,Sangid2010}. Moreover, other microstructural items such as precipitates, non-metallic inclusions and defects, responsible of stochastic response in fatigue, can be explicitly considered \citep{ZHANG2015328}. 

The use of computational homogenization implies solving a boundary value problem on a complex RVE and the Finite Element method (FEM) is the most common choice for microstructure sensitive fatigue models. There are several reasons for this election: the availability of many FEM codes both commercial or open source, the efficient implicit integration of the non-linear problems,   and the accurate representation of complex geometries (including smooth interfaces for grain boundaries) achieved by the use of adaptive meshing. Nevertheless, FEM applied to microstructure based fatigue models presents also some limitations such as the meshing requirement or the high computational cost of the simulations. Due to the stochastic nature of fatigue, tens of RVEs need to be generated, meshed and analyzed. However the generation of quality meshes for a polycrystalline RVEs is not easy to automatize, and a common alternative is to use structured voxel models \citep{Castelluccio2014,Cruzado2017}. Finite element voxel models can be automatically generated but the most interesting features of FE, the adaptation of the mesh to the actual geometry and smooth representation of grain boundaries, is lost when using this type of discretization. The second limitation is the computational cost. The models required for an accurate prediction of fatigue life based on the microfields are particularly large and, in addition, the simulation of several cycles is often required \citep{Cruzado2018}. The FEM in its implicit form, scales by $O(n^2 \text{-} n^3)$ depending on the particular conditions and solver used and this scaling prevents the use of very big meshes or requieres the use of largely parallel codes running in computer clusters.

An alternative to FEM that have become very popular in the last years for solving many homogenization problems are the methods based on the Fast Fourier Transform (FFT) or the spectral solvers.  The original approach was first introduced by P. Suquet and H. Moulinec \citep{Moulinec1998} and the seminal idea consists in solving the non-homogeneous Poisson equation of the equilibrium of microfields in an heterogeneous medium by using a reference material and the Green's functions method. The periodic Green's function in the reference medium for a periodic domain can be easily obtained by transforming the equations to the Fourier space and using the fast Fourier transform (FFT) algorithm to compute the discrete Fourier transform of the fields. FFT based homogenization has become very popular due to its excellent numerical performance (the algorithm scales by $n \log n$) that allows the use of very detailed RVEs with a reduction in computational cost compared to FE, being able to speed up simulations by orders of magnitude \citep{Eisenlohr2013}. For this reason, since its introduction, the framework has been successfully applied for homogenizing the mechanical response of composites \citep{KABEL2015168,MONCHIET2013276}, voided materials \citep{LEBENSOHN20136918,LEBENSOHN20123838} and polycrystalline metals \citep{Lebensohn2001,LEBENSOHN20045347,Eisenlohr2013} under monotonic loading.

However, the use of FFT in computational microstructure sensitive fatigue models remains almost unexplored. The first reason is a practical reason. FEM is a mature technique that allows the use of general constitutive models and most of the researchers in the micromechanics fatigue modeling community have implemented their crystal plasticity models for FEM codes.  Secondly, from a more fundamental viewpoint, FFT solvers might present local fluctuations in the microscopic fields (Gibbs phenomena) that, although they do not play an important role for the overall behavior \citep{Willot2015}, can be critical for predicting fatigue life, since this prediction is based on extreme values of microscopic fields and  not on the homogenized RVE response. The only previous study up to date, according to the authors knowledge, of applying FFT solvers to fatigue prediction was performed by Rovinelli, Sangid and Lebensohn \citep{Sangid2015} where microfields were computed near  real microscopic cracks to extract fatigue  indicator parameters and estimate propagation directions. However, this work was performed under small strains approach and is focused on studying the microfield distribution around existing cracks while the localization of plastic deformation during the cyclic loading (driving force for crack nucleation) of undamaged polycrystals was not explored.  

The objective of this work is to develop a FFT based framework for microstructure sensitive fatigue life prediction and to assess its accuracy respect FEM  simulations. The framework will include (1) a robust implementation of the FFT solver using finite strains and implicit integration based on the variational approach proposed in \cite{Geers2016,Geers2017}, (2) direct use of constitutive equations developed for finite element codes,  including the material consistent tangents, in particular allowing to use any general Abaqus \emph{umat}s subroutine, (3) correction of Gibbs phenomena by the development of new projection operators based on discrete derivatives \citep{Willot2015} allowing to include hard particles or voids in the simulations. To assess the accuracy and efficiency of the framework, simulations of the cyclic response of identical RVEs  will be done with FEM and FFT and the averaged behavior, local microfields and life estimation will be compared.

The paper is arranged as follows. In section 2 the FFT framework for micromechanics based fatigue modeling will be presented. The computational microstructure sensitive fatigue prediction technique is introduced in section 3. Section 4 will show the results of the comparisons between FFT and FEM and section 5 will analyze the computational performance of both codes. Finally, the concluding ideas are summarized in section 6.

\section{FFT framework for micromechanics based fatigue simulations}
A FFT homogenization scheme is adapted here to be used for microstructure based fatigue life prediction of polycrystals. The framework allows the direct use of any crystal plasticity model developed for finite elements and includes the use of discrete projection operators for an accurate resolution of the microfields used to define for fatigue indicator parameters. The FFT homogenization basis is the variational approach to FFT recently developed \citep{Vondrejc2010,Geers2016,Geers2017} as an alternative to classic schemes introduced in \citep{Moulinec1998} that are the common framework used in all the models and codes devoted to polycrystalline homogenization \citep{Lebensohn2001,LEBENSOHN20045347,Eisenlohr2013} previous to this work. This approach is based on the variational concepts underlying FEM  method and introduces several advantages respect to classical schemes, a robust theoretical background, the lack of a reference medium, and a non-linear extension formally identical to FEM that uses Newton method with material tangent matrices.  All these characteristics result in a very efficient alternative. From one side, the linear solver does not requiere a reference medium which properties should be redefined every increment to obtain optimal convergency and the result is as fast as the optimal classical schemes (i.e. accelerated scheme \citep{Michel1999}). From the other side, the non-linear extension permits to reach quadratic convergence under the same conditions as finite element,  allowing the use of large strain increments. A summary of the main ideas and equations of the variational approach proposed in \citep{Geers2016,Geers2017} will be presented below, explicitly developing the treatment of materials with internal variables in the formulation.

\subsection{Variational approach}
The objective of the method is finding the equilibrium deformation gradient and stress microfields in a periodic domain $\Omega_0$ representative of the microstructure of an heterogeneous material. The domain $\Omega_0$  in the reference configuration corresponds to a rectangular parallelepiped characterized by three dimensions, namely $L_1, L_2$ and $L_3$. The domain contains different phases, each one having a particular non-linear behavior, and each point of the domain $\mathbf{x} \in \Omega_0$ belongs to one of those phases. The starting point of the method is the linear momentum balance in the domain $\Omega_0$ under periodic boundary conditions. 
\begin{equation}
\left\{
\begin{array}{c} 
\mathrm{Div} \ \mathbf{P}(\mathbf{x})=\mathbf{0} \\
\mathbf{F}(\mathbf{x})=\overline{\mathbf{F}}+\tilde{\mathbf{F}}(\mathbf{x})\\
\mathrm{with} \ <\mathbf{F(x)}>_\Omega=\overline{\mathbf{F}}\\
\mathrm{and} \ \tilde{\mathbf{F}}(\mathbf{x}) \ \mathrm{periodic}
\end{array}\right.
\label{linear_momentum}
\end{equation}
being ``$\mathrm{Div}$"  the divergence in the reference configuration,  $\mathbf{P}(\mathbf{x})$  the microscopic first Piola-Kirchhoff stress tensor and $\mathbf{F}(\mathbf{x})$ the microscopic deformation gradient. This field is decomposed in its average value in the domain $\overline{\mathbf{F}}$ (macroscopic or far-field   deformation gradient) and the fluctuation field $\tilde{\mathbf{F}}(\mathbf{x})$. The stress at each point of the domain is given by a non-linear constitutive law that models the behavior of the material at that point,
\begin{equation}
\label{eq:constitutive}
\mathbf{P}(\mathbf{x})=\mathbf{P}(\mathbf{F}(\mathbf{x}),\boldsymbol{\alpha}(\mathbf{x}))
\end{equation}
where $\boldsymbol{\alpha}$ is a vector containing the internal variables of the constitutive model accounting for the effect of its deformation history. The deformation gradient solution of the strong formulation of equilibrium  (eq. \ref{linear_momentum}) must be a periodic and curl-free field. A weak formulation of the equilibrium is derived then using the virtual work principle. The problem statement consist in finding for every time of a given macroscopic deformation gradient history $\overline{\mathbf{F}}(t)$ the deformation gradient field $\mathbf{F}(\mathbf{x})$ that fulfills
\begin{equation}\label{weakform}
\int_{\Omega_0}\delta\mathbf{F}(\mathbf{x}):\mathbf{P}(\overline{\mathbf{F}}+\tilde{\mathbf{F}}(\mathbf{x}))\mathrm{d}\Omega_0=0
\end{equation}
being $\delta\mathbf{F}(\mathbf{x})$ any periodic deformation gradient virtual field fulfilling compatibility and periodicity conditions.  All the fields in (eq. \ref{weakform}) are defined at time $t$, although this is omitted to relax the formulation.  Note that the terms related to the work done by external stress are not included because they vanish due to the periodic boundary conditions. 
The compatibility of the virtual field $\delta\mathbf{F}(\mathbf{x})$ is imposed using a projection operator $\mathbb{G}$, which is a fourth order rank linear operator that maps any arbitrary second order tensor periodic field $\boldsymbol{\zeta}(\mathbf{x})$ in its compatible (curl-free) part
\begin{equation}\label{arbitrary}
\delta\mathbf{F}(\mathbf{x}) = (\mathbb{G} \ast \boldsymbol{\zeta})(\mathbf{x})
\end{equation}
where $\ast$ stands for the convolution. The projection operator $\mathbb{G}$ is equivalent to the Green's function of a reference media $\mathbb{\Gamma}$ introduced in classical approaches \citep{Moulinec1998} but here no choice of the reference properties is needed. The standard expression of the projection operator and the modified discrete operator version are given in section \ref{operator}.  Substituting equation (\ref{arbitrary}) into the weak formulation of equilibrium (eq. \ref{weakform}) and exploiting the symmetries of $\mathbb{G}$ leads to the next integral equation

\begin{equation}\label{weakarbi}
\int_{\Omega_0}\boldsymbol{\zeta}(\mathbf{x}):\left[ (\mathbb{G}\ast \mathbf{P})(\mathbf{x})\right] \mathrm{d}\Omega_0=0
\end{equation}
that should be fulfilled for every tensor  test function $\boldsymbol{\zeta}(\mathbf{x})$ belonging to the space of all square-integrable arbitrary tensor fields. Then, the domain $\Omega_0$ is discretized in a voxelized regular grid containing $n_x\times n_y\times n_z$ voxels, being each voxel labeled by three integers $x,y,z$ with $1 \le x \le n_x$; $1 \le y \le n_y$ and $1 \le z \le n_z$. The fields $\mathbf{F}(\mathbf{x})$, $\boldsymbol{\zeta}(\mathbf{x})$ and the internal variables $\boldsymbol{\alpha}(\mathbf{x})$ are approximated by  $\mathbf{F}^h(\mathbf{x})$, $\boldsymbol{\alpha}(\mathbf{x})^h$ and $\boldsymbol{\zeta}^h(\mathbf{x})$ respectively. These approximate fields are obtained by interpolating the values at the center of each voxel using the fundamental trigonometric polynomials as shape functions \cite{Geers2016},  defined by the discrete Fourier's transform.  As an example, the approximation of the deformation gradient $\mathbf{F}^h(\mathbf{x})$ is obtained as
\begin{equation}
\mathbf{F}^h(\mathbf{x})=\sum_{x,y,z=\mathbf{1}}^{n_x,n_y,n_z} N_{xyz}(\mathbf{x}) \mathbf{F}_{xyz} 
\label{interpol}
\end{equation}
where $N_{xyz}(\mathbf{x})$ are the trigonometrical polynomials for the voxel defined by the integers $x,y,z$ and $\mathbf{F}_{xyz}$ is the value of the deformation gradient at that voxel, $\mathbf{F}_{xyz}=\mathbf{F}(\mathbf{x}(x,y,z))$.  The use of the approximate fields in the expression of the virtual work leads to 

\begin{equation}\label{weakarbi2}
\int_{\Omega_0}  \boldsymbol{\zeta}^h(\mathbf{x}) :\left[  \mathbb{G} \ast  \mathbf{P} (\mathbf{F}^h(\mathbf{x}),\boldsymbol{\alpha}^h(\mathbf{x}))\right] \mathrm{d}\Omega_0=  0
\end{equation}
This integral (eq. \ref{weakarbi2}) is computed using the trapezoidal rule, expressing the trigonometrical polynomials in terms of the discrete Fourier coefficients and using the discrete Fourier transform to perform the convolution operation \cite{Geers2016}. Note that, opposed to standard finite elements, the integration points are the center of the voxels, the same positions defining the approximated fields (eq. \ref{interpol}). 
The result of the integral is a sum over the voxels given by
\begin{equation}
\sum_{x,y,z=\mathbf{1}}^{n_x,n_y,n_z}  \boldsymbol{\zeta}_{xyz} : \left[  \mathbb{G} \ast  \mathbf{P} (\mathbf{F}_{xyz},\boldsymbol{\alpha}_{xyz})\right]_{xyz}=  0
\label{discrete_virtual}
\end{equation}
The equation (\ref{discrete_virtual}) must be fulfilled for any arbitrary discrete voxel field $\boldsymbol{\zeta}_{xyz}$, implying that the right hand side of the equation has to be zero at every voxel. If the projection operator is defined in the Fourier space and the convolution is also  performed as a multiplication in the Fourier space, the conditions of weak equilibrium can be written as
\begin{equation}
\mathcal{F}^{-1}\left\{ \hat{\mathbb{G}}:\mathcal{F} \left(\mathbf{P}_{xyz} \right) \right\}=\mathbf{0}_{xyz}
\label{eql_discrete1}
\end{equation}
where $\mathcal{F}$ and $\mathcal{F}^{-1}$ are the discrete Fourier transform and its inverse respectively, $\mathbf{P}_{xyz}$ stands for the first Piola-Kirchhoff  stress at the center of each voxel $\mathbf{P}(\mathbf{F}_{xyz},\boldsymbol{\alpha}_{xyz})$ and $\hat{\mathbb{G}}$ is the value of the projection operator in the discrete Fourier space. The different expressions  used for the projection operator are given in \ref{operator}. Equation (\ref{eql_discrete1}) is an algebraic  system of $9\cdot n_x \cdot n_y \cdot n_z$ equations and in which the unknown is the value of the deformation gradient at each voxel $\mathbf{F}_{xyz}$. This equation can be written in a more simple way as
 \begin{equation}
\mathcal{G}(\mathbf{P}_{xyz}(\mathbf{F}_{xyz},\boldsymbol{\alpha}_{xyz}))=\mathbf{0}_{xyz}
\label{eql_discrete2}
\end{equation}
where $\mathcal{G}$ is a linear map on the vector space in which the voxel fields are defined ($\mathbb{R}^{3\times3\times nx \times ny \times nz}$) acting on a voxel tensor field $\mathbf{A}_{xyz}$ to create a new voxel tensor field $\mathbf{B}_{xyz}$  into the same space and is defined as 
\begin{equation}
\mathcal{G}(\mathbf{A}_{xyz}):=\mathcal{F}^{-1}\left\{\hat{\mathbb{G}}:\mathcal{F}\left(\mathbf{A}_{xyz}\right) \right\}=\mathbf{B}_{xyz}.
\label{eql_discrete3}
\end{equation}

In the case of non-linear material behavior the discrete expression of the weak form of equilibrium given by equation (\ref{eql_discrete2}) defines a non-linear system of algebraic  equations. Moreover, in the case of a material with internal variables, a history of the macroscopic deformation gradient $\overline{\mathbf{F}}(t)$ has to be defined, the time is discretized in increments and a non-linear problem has to be solved for every time increment $t=t_k$.  Let $\overline{\mathbf{F}}(t_k)$ be the value of the macroscopic deformation gradient at time $t_{k}$, the objective is finding the deformation gradient at the voxels $\mathbf{F}^{t_k}_{xyz}$ solving the equation (\ref{eql_discrete2}) and being the deformation gradient $\mathbf{F}_{xyz}^{t_{k-1}}$ and the internal variables $\boldsymbol{\alpha}_{xyz}^{t_{k-1}}$ known at the previous time step. This equation can be solved iteratively by the Newton-Raphson method. To this aim, the deformation gradient field is obtained in an iterative manner, and its value for iteration $i$ is given by
\begin{equation}
\mathbf{F}_{xyz}^{i}=\mathbf{F}_{xyz}^{i-1} +{\delta}\mathbf{F}_{xyz}.
\label{F_dif}
\end{equation}
The stress field is linearized around the deformation gradient
\begin{equation}
\begin{aligned}
\mathbf{P}_{xyz}^{i} & =\mathbf{P}_{xyz}^{i-1}+\delta\mathbf{P}_{xyz}=\mathbf{P}_{xyz}^{i-1}+\left. \frac{\partial \mathbf{P}}{\partial \mathbf{F}}\right|_{(\mathbf{F}=\mathbf{F}_{xyz}^{i-1},\boldsymbol{\alpha}_{xyz}^{i-1})}:\delta \mathbf{F}_{xyz} = \\
& =\mathbf{P}_{xyz}^{i-1}+\mathbb{K}^{i-1}_{xyz}:\delta \mathbf{F}_{xyz}^{i-1}
\end{aligned}
\label{P_dif}
\end{equation}
where $\mathbb{K}$ is the (non-symmetric) material tangent. Combining the equilibrium equation (\ref{eql_discrete2}) with the linearization of deformation gradient and stress (eqs. \ref{F_dif} and \ref{P_dif}) the next equation is derived for the iteration $i$
\begin{equation}\label{linequil}
\mathcal{G}( \mathbb{K}^{i-1}_{xyz}:\delta \mathbf{F}_{xyz})=-\mathcal{G}(\mathbf{P}(\mathbf{F}_{xyz}^{i-1}))
\end{equation}
If a new linear map is defined, the resulting expression can be written as
\begin{equation}\label{linequil2}
\mathcal{G}_{\mathbb{K}^{i-1}}(\delta \mathbf{F}_{xyz})=-\mathcal{G}(\mathbf{P}(\mathbf{F}_{xyz}^{i-1}))
\end{equation}
where the new linear operator $\mathcal{G}_{\mathbb{K}^{i-1}}$ is defined as
\begin{equation}
\mathcal{G}_{\mathbb{K}^{i-1}}(\mathbf{A}_{xyz}):=\mathcal{G}(\mathbb{K}^{i-1}_{xyz}: \mathbf{A}_{xyz})
\label{newlinop}
\end{equation}
The expression in equation \ref{linequil2} is a linear system of equations in which the unknown is the correction at the iteration $i$ of the deformation gradient $\delta \mathbf{F}_{xyz}$. In this system, the coefficient matrix is very large (dimension $(9\ nx\ ny \ nz)^2$) and dense, but it is not necessary to explicitly compute it because it is fully defined by the linear operator $\mathcal{G}_{\mathbb{K}^{i-1}}$. This type of equation can therefore be solved using  a Krylov iterative solver  where, instead of computing and storing the coefficient matrix, a linear operator can be used instead. The conjugate gradient method \cite{Stiefel1952} has been chosen to solve the system due to its good performance for this type of systems.  Moreover, the system in eq. (\ref{linequil2}) is indefinite so the use of an iterative descent method is mandatory because, contrary to direct solvers, these methods allow to obtain a solution without eliminating any equation \citep{KAASSCHIETER1988265}. This deficiency in range of the operator $\mathcal{G}_{\mathbb{K}^{i-1}}$ is due to its symmetries and to the null value of both the zero and Nyquist frequencies.

For a given time increment $t_k$ corresponding to a macroscopic deformation gradient $\overline{\mathbf{F}}(t_k)$, the solution of the linear equation (\ref{linequil2}) provides a new Newton correction of the deformation gradient, and the equilibrium is reached when the right-hand side of the equation $\mathcal{G}(\mathbf{P}_{xyz})$ is sufficiently small. As in any finite element framework, once reached the equilibrium for time $t_k$, the internal variables are updated and new time increments are solved until reaching the final time.

\subsection{Standard and discrete versions of the projection operator}
\label{operator}
The results obtained by FFT might present the well-known Gibbs oscillation phenomena in microstructures with a large contrast between phases.  In the case of polycrystals, this contrast appears due to the differences in the elasto-plastic response of adjacent grains with different orientations and also due to the presence of non-metallic phases.  To avoid this undesirable oscillations, a very interesting approach consist in replacing  the continuum derivative of fields in the Fourier space with a discrete derivative, as proposed by Willot \citep{Willot2014}.   The derivative  of a periodic function $f$ in $L$ using Fourier transforms is expressed as
\begin{equation}
\frac{\mathrm{d}}{\mathrm{d}x}f(x)=\mathcal{F}^{-1}\left(\mathrm{i} (q/L) \mathcal{F}\left(f(x)\right)\right)
\label{eq:1Ddif}
\end{equation}
where $q$ are the Fourier frequencies and $\mathrm{i}$  represents the imaginary unit. If the function is discretized in $n$ voxels in the real space, $q$ corresponds to
\begin{equation}
q^{(i)}= \left\{\begin{array}{c} 2\pi \frac{n/2-i}{n}  \ \text{if n even} \\2\pi \frac{(n+1)/2-i}{n} \ \text{if n odd} \end{array} \right. \ , \ i=1,\ldots,n
\label{eq:xis}
\end{equation}

A discrete version of the  derivative can be obtained transforming to the Fourier space the approximation of the continuum derivative in the real space as the finite difference between adjacent points . Comparing the Fourier transform of the discrete derivative with the standard differentiation operator, a discrete  differentiation operator  can be obtained by substituting in eq.(\ref{eq:1Ddif}) the frequencies $q$ by some modified frequencies $q'=q'(q)$, which expressions depend on the type of discrete differentiation performed in the real space (centered scheme, forward or backward differences, etc)  \citep{Willot2014}. Using this discrete expression of the  derivative , a discrete projection operator will be obtained next.

The standard projection operator in the Fourier space introduced in equation (\ref{eql_discrete1}), $\hat{\mathbb{G}}$, is a fourth order rank tensor given for every frequency vector $\mathbf{q}=(q_1,q_2,q_3)$ by
\begin{equation}
\hat{G}_{ijkl} = \left\{ \begin{array}{c} 0_{ijkl} \quad \text{for null and Nyquist frequencies} \\ \delta_{ik} \frac{q_j q_l L_m^2}{q_m q_m L_j L_l} \quad \text{for the rest of frequencies} \end{array} \right.
\label{eq:G}
\end{equation}
where the vector $\textbf{L}$ contains the initial length of the domain $\Omega_0$ in each direction, $(L_1,L_2,L_3)$. The componentes of the vector $\mathbf{q}$ are defined as $q_1=q^{(a)},q_2=q^{(b)},q_3=q^{(c)}$ where each component is defined by one of the frequencies of the corresponding direction, and is determined by $a=1,\ldots,n_x$,  $b=1,\ldots,n_y$ and $c=1,\ldots,n_z$  using equation (\ref{eq:xis}).

The discrete operator is obtained by replacing in equation (\ref{eq:G}) each vector $\mathbf{q}$ by its modified counterpart $\mathbf{q}'$ with $q'_1=q'^{(a)},q'_2=q'^{(b)},q'_3=q'^{(c)}$. In this study we have used the rotated centered difference for obtaining the discrete derivatives in the real space \cite{Willot2014}. This scheme performs the finite differentiation  using the value of the tensor fields at the center of the voxels and the value of the displacement field at the voxel corners \citep{Willot2015}.   The resulting modified wave vectors are given by
\begin{equation}
\begin{array}{c}
q'^{(a)} = \frac{1}{4} \tan{\frac{q^{(a)}}{2}} \left( 1 + e^{\mathrm{i}q^{(a)}} \right) \left( 1 + e^{\mathrm{i}q^{(b)}} \right) \left( 1 + e^{\mathrm{i}q^{(c)}} \right)\\
q'^{(b)} = \frac{1}{4} \tan{\frac{q^{(b)}}{2}}\left( 1 + e^{\mathrm{i}q^{(a)}} \right) \left( 1 + e^{\mathrm{i}q^{(b)}} \right) \left( 1 + e^{\mathrm{i}q^{(c)}} \right)\\
q'^{(c)} = \frac{1}{4} \tan{\frac{q^{(c)}}{2}} \left( 1 + e^{\mathrm{i}q^{(a)}} \right) \left( 1 + e^{\mathrm{i}q^{(b)}} \right) \left( 1 + e^{\mathrm{i}q^{(c)}} \right)
\end{array}
\end{equation}
and replacing these new vectors in the original expression of the operator (\ref{eq:G}), the resulting modified projection operator, $\mathbb{\hat{G}}_{ijkl}'$ for every frequency vector $\mathbf{q}$ can be expressed as
\begin{equation}
\hat{G}_{ijkl}' = \left\{ \begin{array}{c} 0_{ijkl} \quad \text{for null and Nyquist frequencies} \\  \delta_{ik} \frac{\tan\left( \frac{q_j}{2}\right) \tan\left( \frac{q_l}{2}\right) L_m^2}{\tan^2\left( \frac{q_m}{2}\right) L_jL_l} \quad \text{for the rest of frequencies} \end{array} \right.
\end{equation}

\subsection{Computational aspects}
The FFT framework presented above has been programmed in a code called \emph{FFTMAD}. The code is mainly written in Python, but including some external routines in Fortran to accelerate the simulations. \emph{FFTMAD} contains different approaches for solving the boundary value problem, the variational approach previously described \cite{Geers2016,Geers2017} and the \emph{classical schemes} including the basic scheme \citep{Moulinec1998}, accelerated scheme and augmented Lagrangian scheme \citep{Michel1999}. The variational approach is implemented for both, small and finite strains. In all the schemes both the standard operators and the modified versions defined from discrete differentiation (section \ref{operator}) can be selected.

Respect non-linear material behavior, FFTMAD includes the possibility of using any general non-linear material (including history dependent materials with internal variable) in both finite and small strain theory by including the corresponding material subroutine. The code admits subroutines for material behavior directly written in a simple Python framework and also includes a wrapper for plugging to the code any Abaqus user material subroutine (\emph{umat}) written in Fortran. This option has been successfully tested for several material subroutines including hyperelastic materials, small strain elasto-plasticty and finite strain crystal (visco-)plasticity models. Finally, the plugin of other format of material subroutines or the link with a material library as \emph{IRIS} \citep{Portilloetal2017} is quite easy with the structure of the program.  Due to the large community of researchers using Abaqus as FEM code, the particular details of the adaptation of constitutive equations defined as a \emph{umat} in the FFT framework are detailed in \ref{tangdev}. 

Other  features  have been implemented in the code for making it specially efficient for large simulations, an automatic time incrementation algorithm and the parallelization of the heaviest operations in the code for a shared memory framework. All the tensor field operations in the discrete voxel or Fourier spaces are programmed in Fortran and parallelized in threads with \textit{openMP}. The evaluation of the constitutive equations is also distributed in threads using the same approach. Finally, the direct and inverse Fourier transforms are computed in parallel by the use of \textit{FFTW} package. In addition to threads based parallelization, a version of GPU parallelization using \textit{pyCUDA} has been implemented for further improvement of the efficiency in the FFT transformation and tensor field operations.

\section{Micromechanics based fatigue modeling}
In this section, the general aspects of microstructure sensitive fatigue modeling will be briefly reviewed. This includes the description of the computational homogenization framework, the crystal plasticity model for the grain behavior and the definition of the fatigue indicator parameters to estimate fatigue life.

\subsection{Polycrystalline homogenization}
The first step towards life prediction using micromechanics based fatigue models is obtaining the distribution of the microfields within the polycrystal microstructure after the cyclic behavior becomes stable (hysteresis loop does not change with new cycles). To this aim, the response of the polycrystal for a given macroscopic cyclic strain history is simulated solving the periodic boundary value problem defined for each time by $\mathbf{F}(t)$ (the averaged value of $\mathbf{F}(t)$) in an RVE of the polycrystal microstructure. This boundary value problem can be solved using either FEM or the FFT framework previously described and using crystal plasticity as constitutive equation of the grains forming the microstructure.

The RVE of a polycrystalline material consists in an aggregate of grains with grain shape, size and orientation distributions representative of the actual microstructure, which can be obtained from experimental measurements \cite{Cruzado2015}. The microstructure for this study is obtained using a modified Voronoi tessellation in which the points defining the grain center and the corresponding weights are defined to fulfil the target grain size distribution and grain spheroidicity. This generation is performed using the open source code  \textit{Neper} \cite{Quey2011}. Finally, the RVE geometry is discretized by rasterizing it in voxels for FFT and in cubic elements for FEM in order to have identical microstructures and discretization for both solvers.

The application of the macroscopic cyclic deformation history  $\mathbf{F}(t)$ to the RVE is direct in FFT.  In the case of FE, periodic boundary conditions are introduced through multipoint constraints between nodes and implemented using Lagrange multipliers. In this framework, the macroscopic deformation gradient is introduced applying a relative displacement between opposite faces of the cube using three different master nodes, one for face pair. More details about the finite element simulations can be found in \citep{HerreraSolaz2014}.  The resulting macroscopic stress is obtained by volume averaging of the microscopic stress field. Simulations were carried out in Abaqus/standard for FEM and FFTMAD for FFT, using in both cases the same RVE, discretization, deformation history, and crystal plasticity model through a \emph{umat} subroutine.

\subsection{Crystal plasticity model}
The  crystal plasticity model recently developed by A. Cruzado, J. LLorca and J. Segurado \cite{Cruzado2017} has been used as constitutive equation for the grains in this study. The model is a phenomenologic cyclic plasticity model and is able to account for the Bauschinger effect, ratcheting and cyclic softening characteristic of many FCC metallic alloys, including polycrystalline superalloys. The model description and implementation is described in detail in \cite{Cruzado2017} but the main features are briefly reviewed here for completeness. 

The crystal plasticity model is elasto-visco-plastic and assumes that plastic deformation occurs by dislocation glide along the 12 octaedrical slip systems. The plastic slip rate in each slip system, $\dot{\gamma}^\alpha$ depends on the resolved shear stress on the slip system, $\tau^\alpha$, following a power-law,
\begin{equation}\label{eq:gamma_dot}
\dot{\gamma}^\alpha=\dot{\gamma}_0 \left(
  \frac{| \tau^\alpha-\chi^\alpha |}{g_c^\alpha+g_s^\alpha}\right)^\frac{1}{m}\mathrm{sign} (\tau^\alpha-\chi^\alpha)  
\end{equation}

\noindent where $\dot{\gamma}_0$ is the reference strain rate, ${m}$ the rate sensitivity exponent. In equation (\ref{eq:gamma_dot}), the critical resolved shear stress on the ${\alpha}$ slip system is the denominator and has two contributions, a monotonic term ${g}_c^\alpha$ that evolves following the Asaro-Needleman model \cite{Asaro1985} and a cyclic softening term ${g}_s^\alpha$, which is negative, and which evolution is dictated by the accumulated cyclic plastic slip. The flow rule (eq. \ref{eq:gamma_dot}) includes a dependency of the slip rate with a back stress, $\chi^\alpha$, that determines the kinematic hardening contribution and which evolution follows the modified version of the Ohno and Wang model proposed in \cite{Cruzado2017}.

\subsection{Fatigue indicator parameters}
When fatigue crack formation and subcritical growth follows crystallographic paths the evolution of the microfields during the simulation of a loading cycle can be used to predict fatigue life. To this aim, microscopic parameters that reflect the local driving forces for fatigue crack formation, named Fatigue Indicators Parameters (FIPs)  \cite{McDowell2010}, are obtained from the microscopic fields obtained in the FFT or FEM simulations . These parameters vary across the RVE depending on the local features of the microstructure and can be related to the number of cycles necessary to nucleate a crack. Different FIPs have been proposed in the literature as driving force to correlate with fatigue life and two of them will be used in this study, the accumulated plastic slip \citep{Manonukul2004, Sweeney2012} and plastic strain energy \citep{Sweeney2014, Wan2014,Cruzado2018}. The local value of the accumulated plastic slip per cycle, $P_{cyc}$ is defined as
\begin{equation}\label{eqplas}
P_{cyc}(\mathbf{x})= \int_{cyc} \mathbf{L_p}(\mathbf{x}) : \mathbf{L_p}(\mathbf{x}) \mathrm{d}t
\end{equation}
\noindent where $\mathbf{L_p}$ is the plastic velocity gradient. The local value of the density of energy dissipated per cycle defines in each slip system defines the plastic strain energy FIP, $W_{cyc}^\alpha$,
\begin{equation}\label{strainenergy}
W_{cyc}^\alpha(\mathbf{x})= \int_{cyc} \tau_\alpha\dot{\gamma}_\alpha(\mathbf{x}) \mathrm{d}t
\end{equation}
\noindent where $\tau_\alpha$ is the resolved shear stress and $\dot{\gamma}_\alpha$ is the shear strain rate on the slip system $\alpha$. The physical meaning of these values is the local accumulation of damage per cycle and it must be highlighted that these parameters are computed in each material point.

The local value of the FIPs may depend on the actual discretization \citep{Sweeney2013} and in order to reduce this dependency and also to introduce explicitly a physical volume in which the nucleation process takes place, non-local measures of the FIPs are defined. These non-local FIP values are the average of the corresponding FIP local value in a given integration volume. The maximum value of the non-local FIP map in the RVE is taken as the parameter representative of the fatigue damage accumulation per cycle in the full microstructure and is used to estimate the number of cycles for crack nucleation. Two integration regions are used in this study, the volume occupied by a grain and the volume defined, for each integration point, by the bands of a given width contained in the grain and parallel to the slip planes \citep{Castelluccio2015}. As a result, two values are used to predict the fatigue life of a given RVE, the grain averaged accumulated plastic slip per cycle, $P_{cyc}^g$
\begin{equation}
P_{cyc}^g=\max_{i=1,ng}  \frac{1}{V^g_i}\int_{V^g_i} P_{cyc}(\mathbf{x})\mathrm{d}V^g_i
\label{Pmax}
\end{equation}
\noindent where $nb$ is the total number of grains and $V_i$ is the volume of the $i$-th grain, and the crystallographic plastic strain energy
\begin{equation}
W_{cyc}^b=\max_{i=1,nb} \left \{\max_{\beta_i} \frac{1}{V^b_i}\int_{V^b_i} W_{cyc}^{\beta_i}(\mathbf{x})\mathrm{d}V^b_i \right \}
\label{Wmax}
\end{equation}
\noindent where $\beta_i$ are the different slips systems contained in the slip plane parallel to the band $i$, $V_i$ is the volume of that band and $nb$ is the total number of bands in the microstructure.

\section{Results and discussion}

In this section, the validity of the FFT framework proposed for micromechanics based fatigue modelling will be assessed. To this aim, the macroscopic response and microscopic fields obtained for different cyclic loading histories using both the FEM and the FFT framework will be compared. The parameters defining the probability for a given RVE to nucleate a crack will be computed using both numerical approaches and the impact in the fatigue life estimation of the differences found between them will be analyzed. Finally, simulations will be made with RVEs containing non-metallic hard inclusions to check the ability of the discrete projection operators proposed to alleviate Gibbs oscillations phenomenon.

\subsection{Material and test description}
The benchmark used for the comparison is the cyclic deformation of a polycrystalline FCC alloy following the constitutive model described in the previous section. The parameters of the crystal plasticity model used for the study are given in Table \ref{Table:CP_parameters} (see \cite{Cruzado2017} for details of the model) and are typical of a polycrystalline superalloy with Bauschinger effect and cyclic softening, such as Inconel 718 alloy.

\begin{table}[H]
\centering
\begin{tabular}[5pt]{lcccc}
\hline\noalign{\vskip2pt}
\multirow{2}{*}{Elastic}
 & C$_{11}$(GPa) & C$_{12}$(GPa) & C$_{44}$(GPa) \\[0.2cm]
 & 233.2 & 160.8 & 98.4  \\
\hline\noalign{\vskip2pt}
\multirow{2}{*}{Viscoplastic}
 & $m$ & $\dot{\gamma}_0$ &  & \\[0.2cm]
 & 0.017 &    2.42 $10^{-3}$ &  & \\[0.5cm]
 \hline\noalign{\vskip2pt}
 \multirow{2}{*}{Isotropic hardening} & $\tau_0$ (MPa) & $\tau_s$ (MPa) & $h_0$ (MPa) & $q_{\alpha\beta}$ \\[0.2cm]
 & 390.5$\tau_0$ & 420.7 & 2039 & 1 \\[0.5cm]
\hline\noalign{\vskip2pt}
\multirow{2}{*}{Kinematic hardening} & $c$ (MPa) & $d$ & $mk$ \\[0.2cm]
 & 23625.4 & 30.3 & 20.5  \\[0.5cm]
\hline\noalign{\vskip2pt}
\multirow{2}{*}{Cyclic softening} & $\tau_s^{cyc}$ (MPa) & $h_{1}$ (MPa) & $h_2$ (MPa) & \\[0.2cm]
 & 29.7 & 25.5 & 0.00122 & \\[0.5cm]
\noalign{\vskip2pt}\hline
\end{tabular}
\caption{Parameters of the crystal plasticity model in \cite{Cruzado2017} for a polycrystalline alloy}
\label{Table:CP_parameters}
\end{table}

\begin{figure}[H]
\includegraphics[width=.32\textwidth]{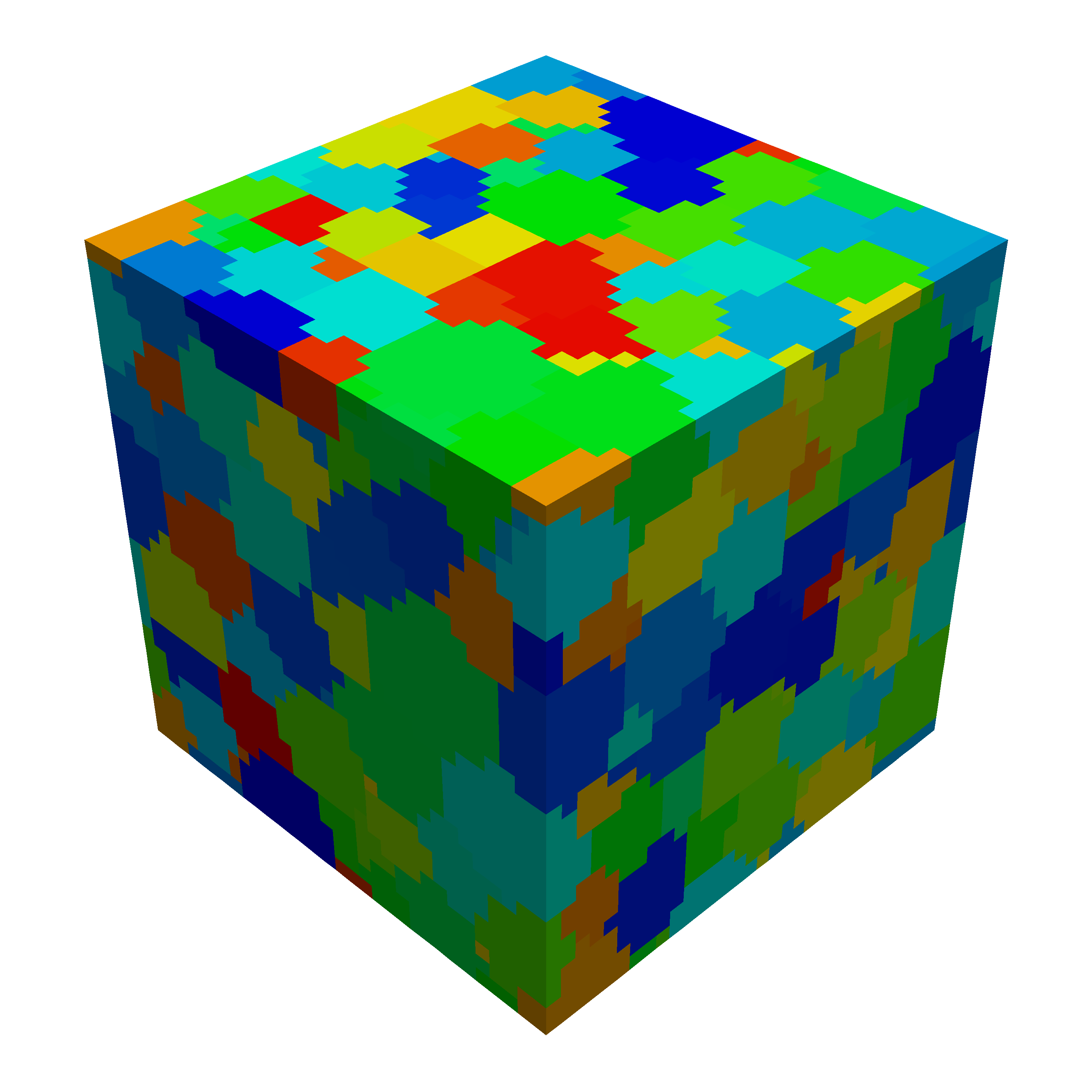} 
\includegraphics[width=.32\textwidth]{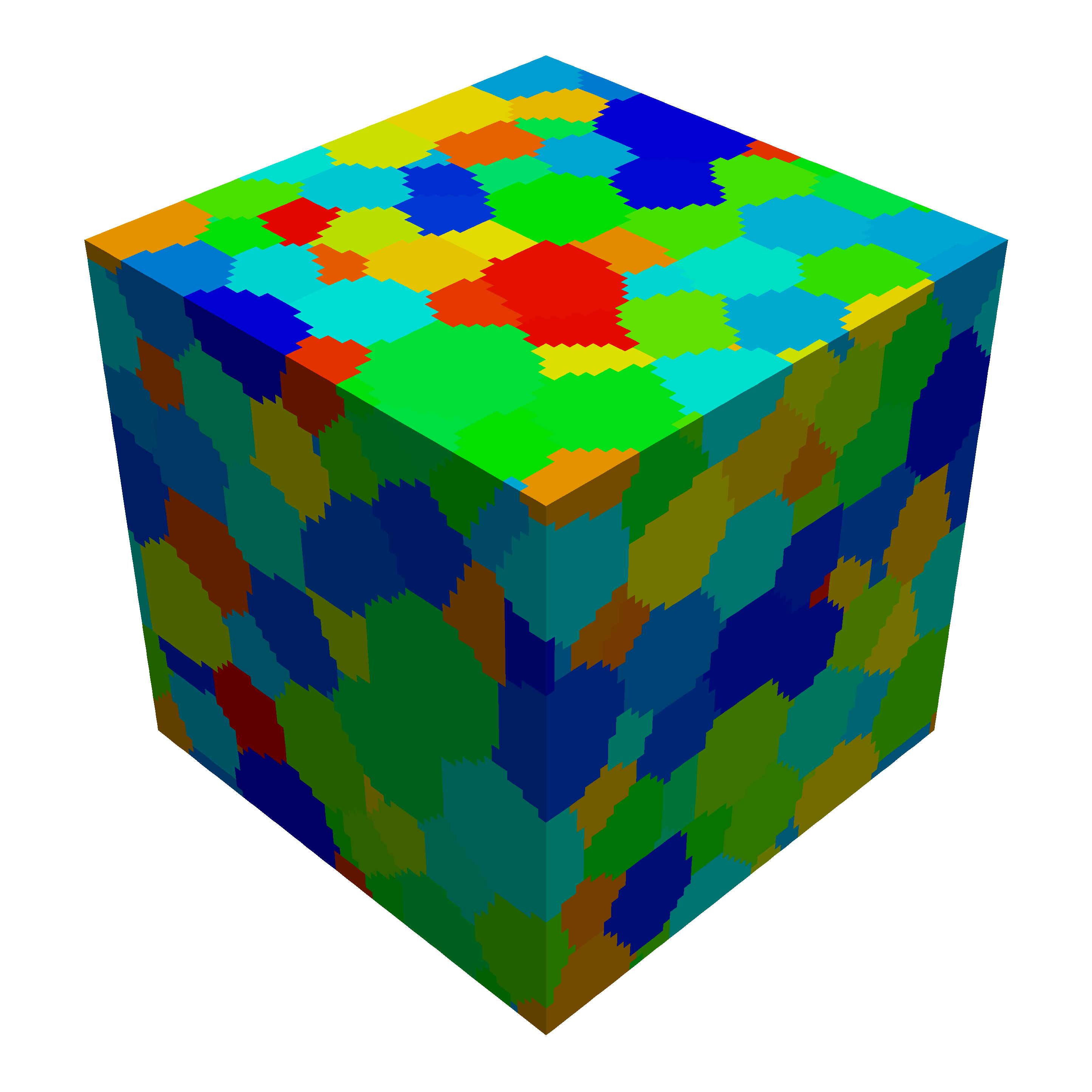} 
\includegraphics[width=.32\textwidth]{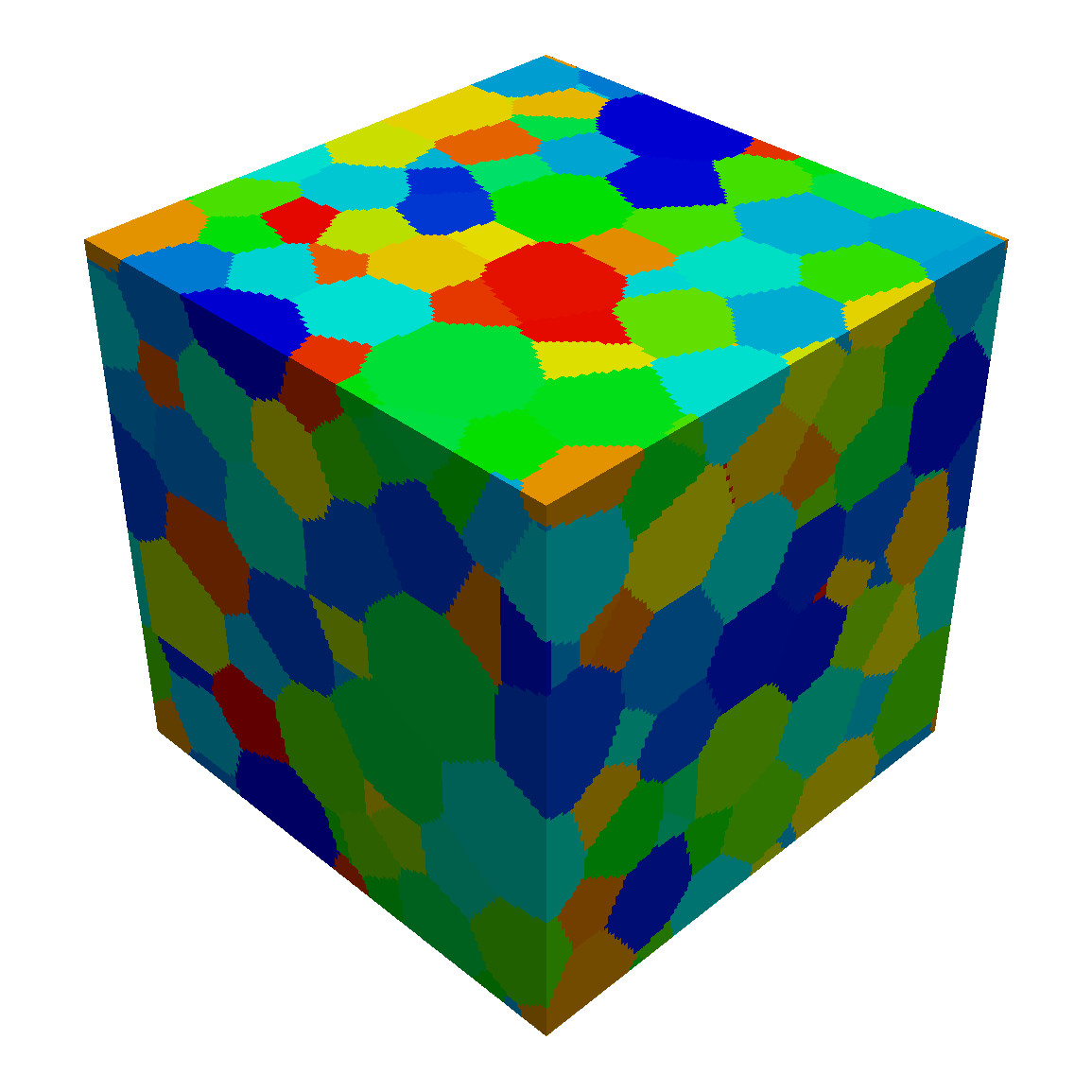} 
\includegraphics[width=.32\textwidth]{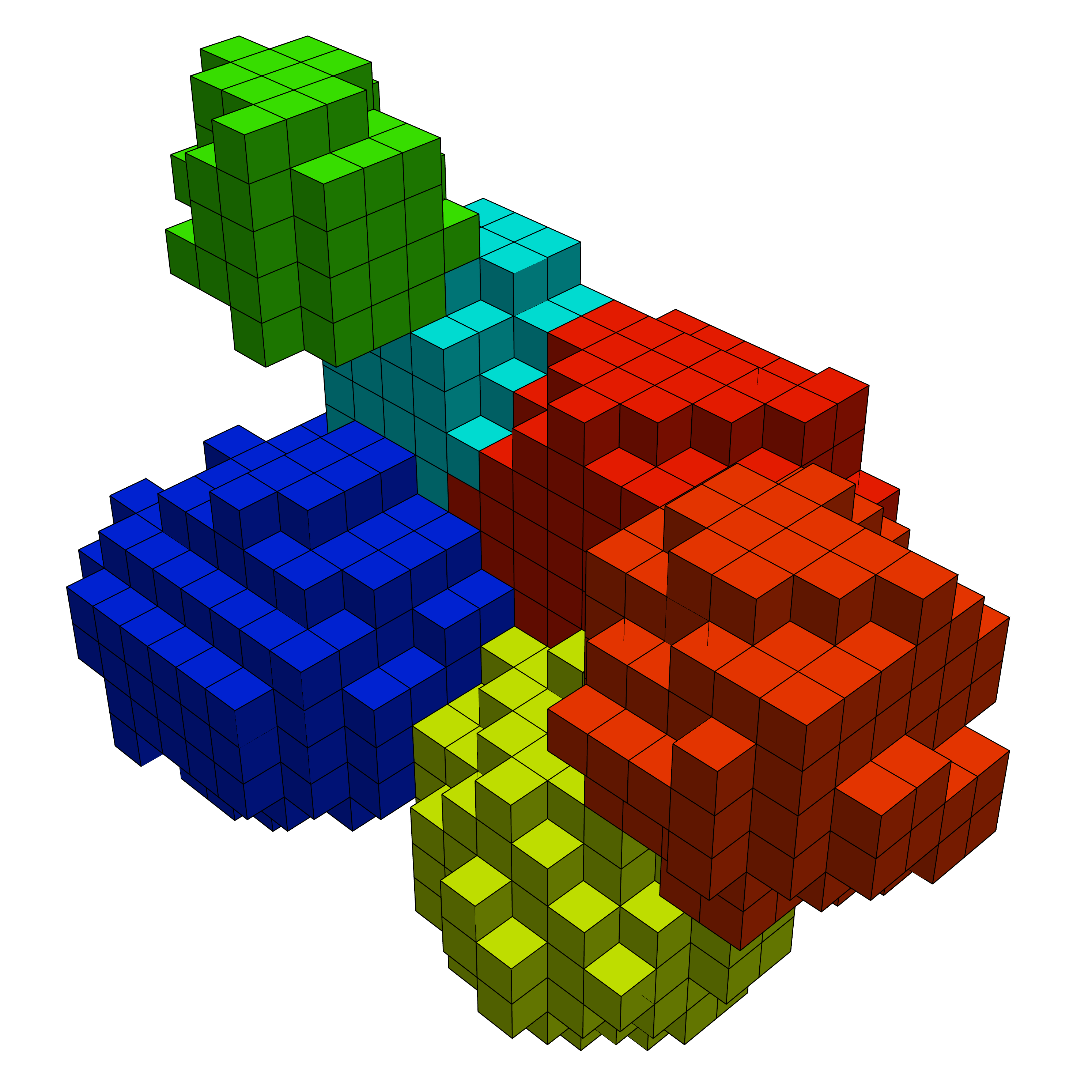} 
\includegraphics[width=.32\textwidth]{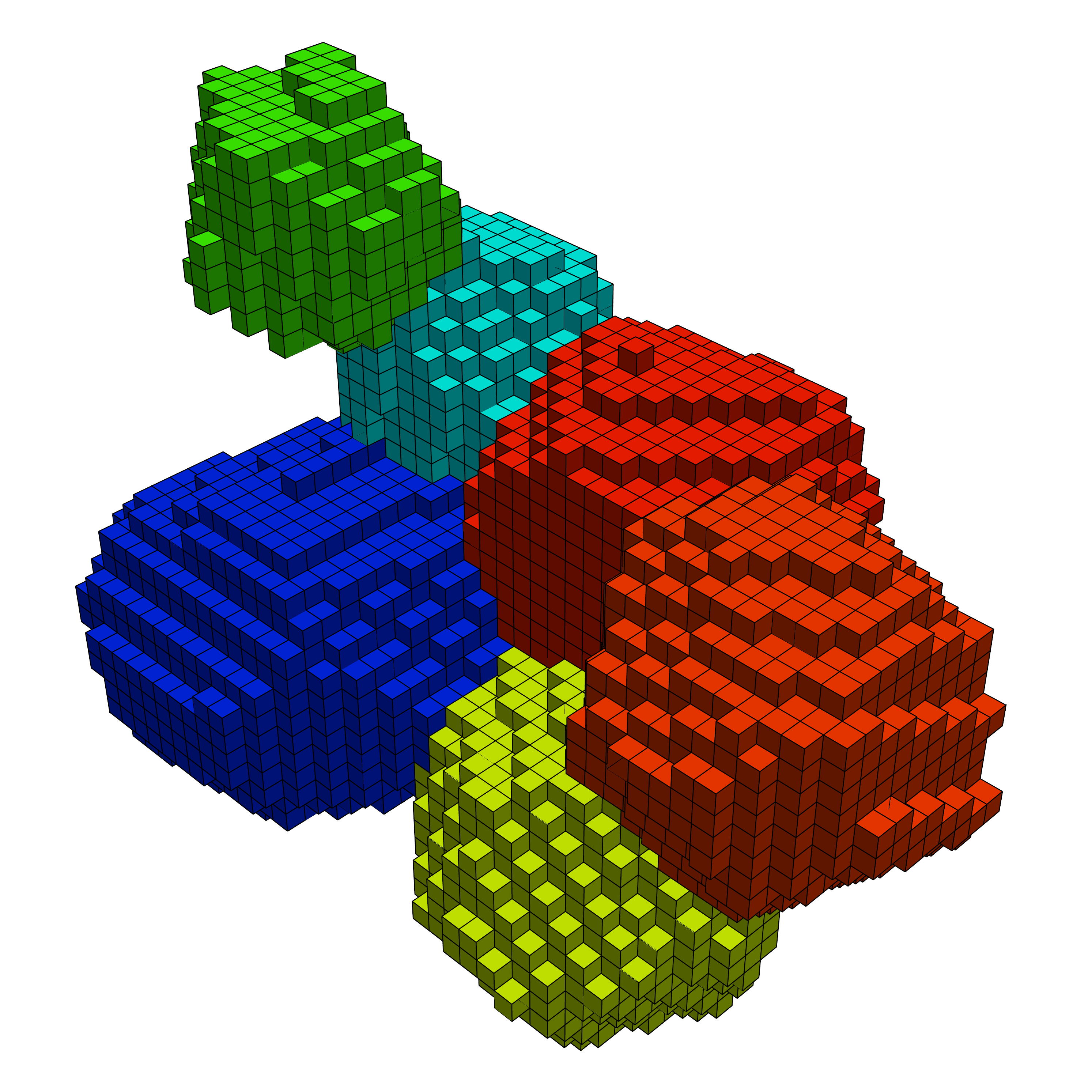} 
\includegraphics[width=.32\textwidth]{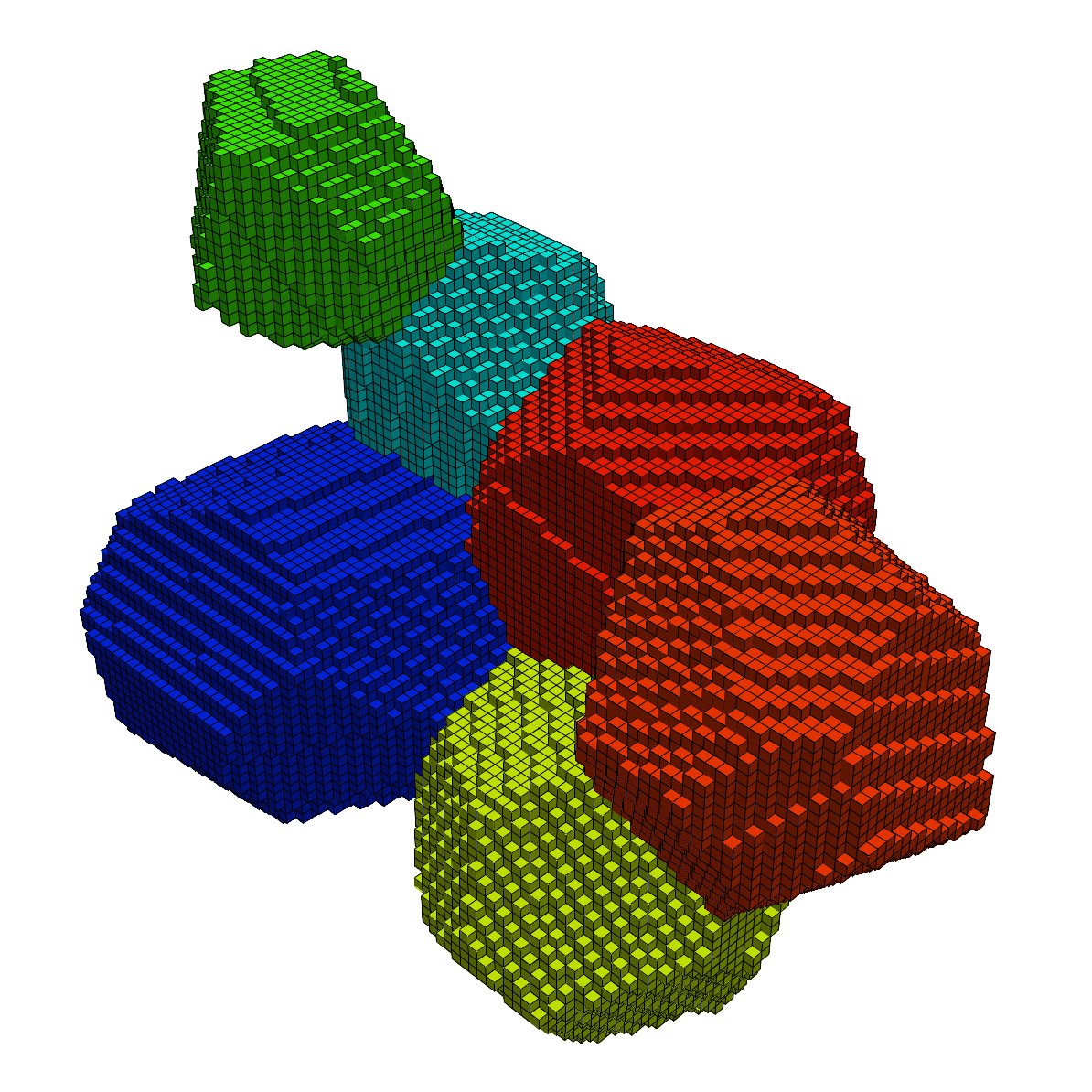} 
\caption{Upper line, discretization of the RVE using $32^3$ voxels, $64^3$ voxels and $128^3$ voxels. Lower line, details of a cluster of grains in the corresponding models}
\label{models}
\end{figure}

The microstructure of the material is represented using a cubic RVE and the grain geometry is generated as a weighted Voronoi tessellation using the open source code \textit{Neper} \cite{Quey2011}. The grain sizes follow a log-normal distribution with a mean value of the grain diameter of $10\mu$m and standard deviation of 0.5. The grains shape is approximately equiaxial by selecting a target spheroidicity equal to 1. Three different models will be considered in this study. All the models share the same geometry ---corresponding to the Voronoi tessellation of a particular arrangement of centers  containing 235 grains--- but are discretized with different number of voxels, $32^3$, $64^3$ and $128^3$, Fig. \ref{models}.

In the case of the finite element simulations, each voxel is used to define a cubic linear element (C3D8 in Abaqus) where Selective Reduced Integration (SRI) method is used to avoid element locking. It must be noted that the complexity of both models is equivalent because the number of independent nodes in the FEM model is identical to the number of voxels of the FFT approach. 

The models are subjected to a nearly uniaxial-stress cyclic strain history. In order to apply this load history using strain control, the applied deformation gradient is given by
\begin{equation}
\mathbf{F}(t)=\left[\begin{array}{ccc}
F(t) & 0 & 0 \\0 &1/\sqrt{F(t)} &0\\ 0& 0 & 1/\sqrt{F(t)} \end{array}\right],
\label{eqFt}
\end{equation}
that is a purely isochoric deformation which, due to the fully incompressible nature of the plastic deformation, results in a stress state very close to uniaxial stress. The value of the function $F(t)$ is set to represent a triangular cycle with a period $T=4$s, macroscopic strain amplitude $\varepsilon_{max}$ and $R_{\varepsilon}=0$. If $t'$ is the time relative to the beginning of the cycle, the deformation gradient in the loading direction is given by

\begin{equation}
F(t')=\left\{ \begin{array}{c} (\frac{1+\varepsilon_{max}}{T})2t' \ if \ $t'$<T/2 \\ 1+2\varepsilon_{max}-(\frac{\varepsilon_{max}}{T}) 2t' \ if \ $t'$>T/2 \end{array} \right.
\label{eqFt2}
\end{equation}

For the FEM simulations, the same deformation history is applied imposing the history of the relative displacement between the corresponding faces in the periodic RVE through the periodic boundary conditions using the master nodes of each face pair \citep{Segurado2013}. In both simulation frameworks the time increments used are the same in order to allow a direct comparison of the results. 

\subsection{Simulation of the cyclic behavior of polycrystals}
 In this section the macroscopic and microscopic response of the polycrystal under two different cyclic loading cases and two different number of cycles will be obtained using the FFT framework. The model results will be compared with the corresponding FEM simulations.

\subsubsection{Macroscopic response}

\begin{figure}[H]
\includegraphics[width=.49\textwidth]{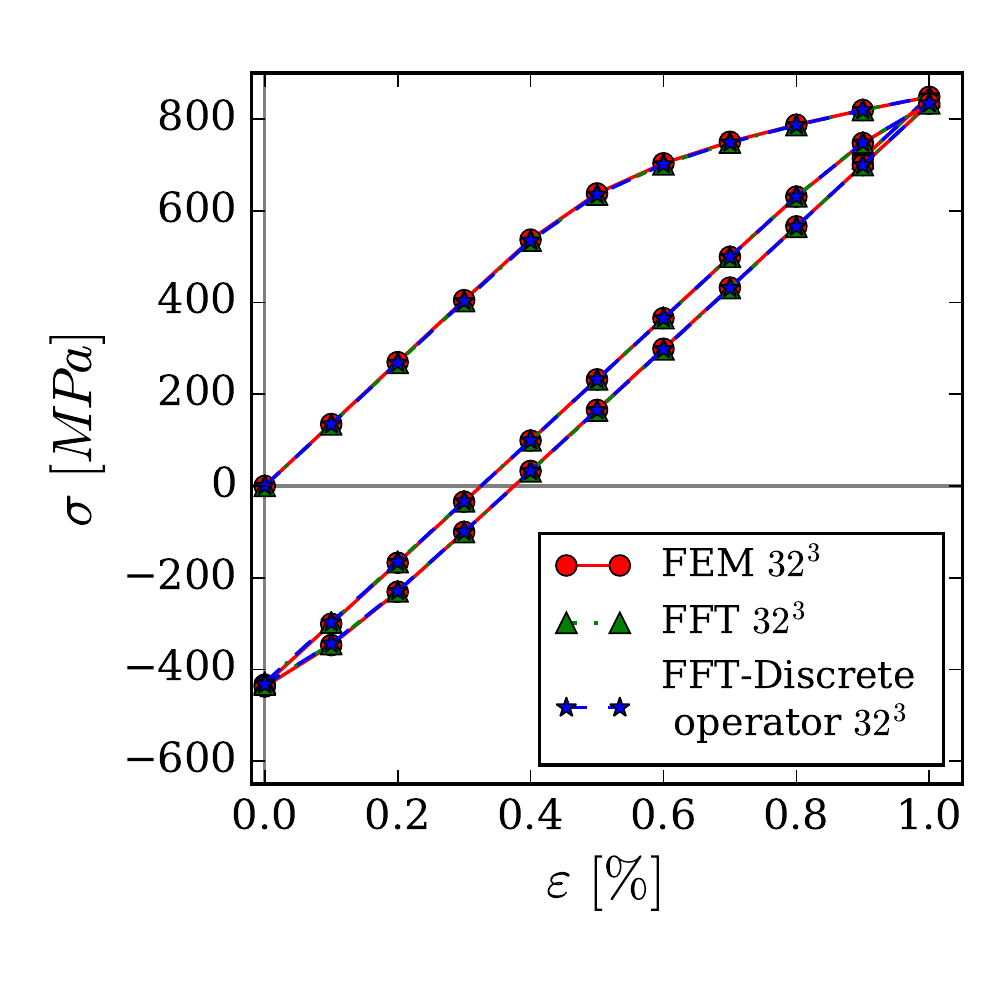}
\includegraphics[width=.49\textwidth]{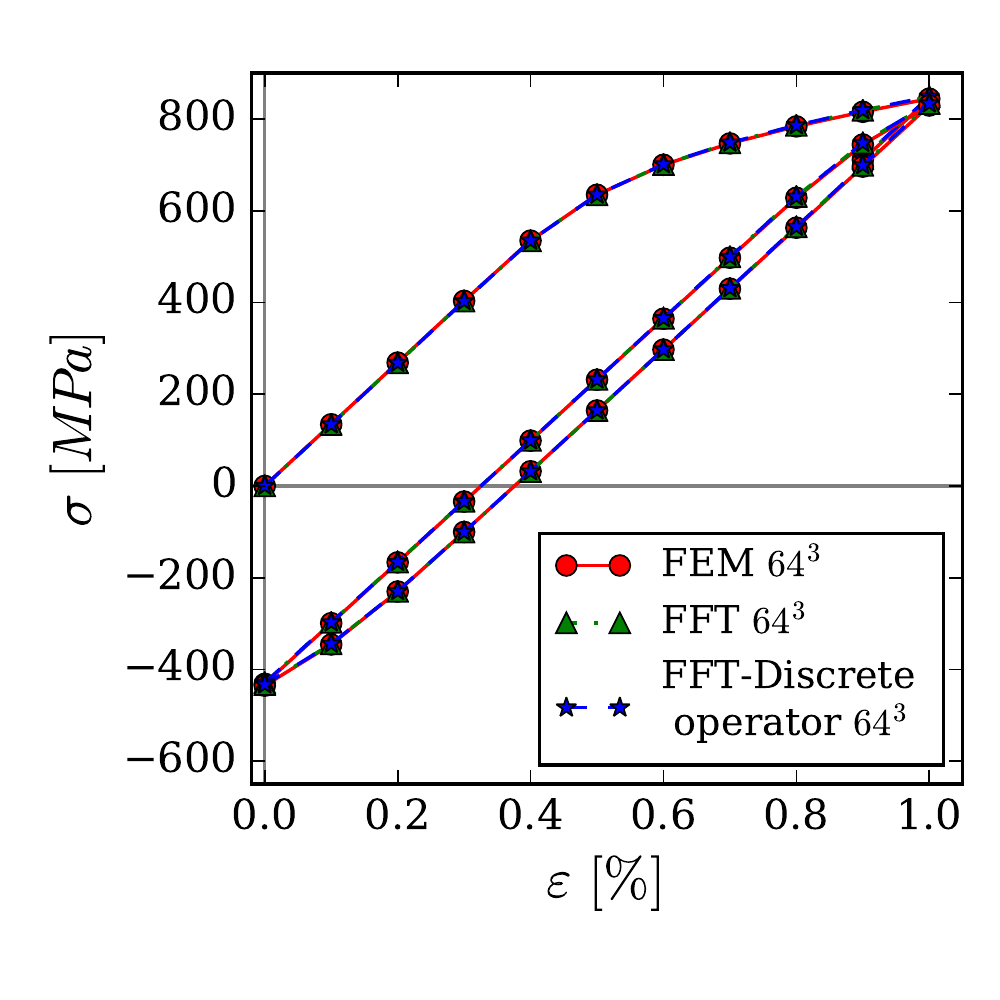}
\caption{Comparative stress-strain curve for a model containing (a) $32^3$ voxel/elements  and (b) $64^3$ voxel/elements}
\label{macro}
\end{figure}

First, a cyclic loading following (eq. \ref{eqFt2}) with macroscopic strain amplitude of  $\varepsilon_{max}=1\%$ is applied to the different discretizations of the polycrystal. This strain range is relatively large for fatigue considerations and the resulting fatigue regime will be Low Cycle Fatigue.  The macroscopic stress-strain response in the loading direction obtained in the numerical simulation of three cycles is represented in Figure \ref{macro} for both FEM and FFT and for two discretizations, $32^3$ (Fig. \ref{macro}(a)) and $64^3$ voxels (Fig. \ref{macro}(b)). In the case of the FFT simulations the results using standard projection operators and the operators based on the discrete derivatives (Section \ref{operator}) are depicted. 

The results in Figure \ref{macro} show that FEM and FFT macroscopic solutions are practically superposed. From a quantitative view point, a relative error of the stress-strain curve predict with FFT respect the FEM solution at time $t$ is defined as
\begin{equation}
Error(t)[\%]=\frac{|\sigma (t)-\sigma_{FEM}(t)|}{max(\sigma_{FEM})}\cdot 100 \text{,}
\label{eq:error}
\end{equation}
and using this definition it was found that the maximum error is below 0.9\% in the case of $32^3$ voxels. When the model is refined to $64^3$ the differences became smaller than 0.5\%.    Finally, it should be noted that the use of projection operators based on discrete derivatives does not influence the macroscopic result of the FFT simulation.

To analyze the influence of the local history in the overall response of the polycrystal two new cyclic simulations are performed. The first case consists in three full cycles with a large macroscopic strain amplitude ($\varepsilon_{max}=5\%$) and the second  one corresponds to 100 cycles with the original value of the macroscopic strain amplitude, $\varepsilon_{max}= 1\%$. In both cases, the differences between FFT, discrete operator FFT and FEM are still very small and the resulting curves are not easy to distinguish.

\begin{figure}[H]
\includegraphics[width=.32\textwidth]{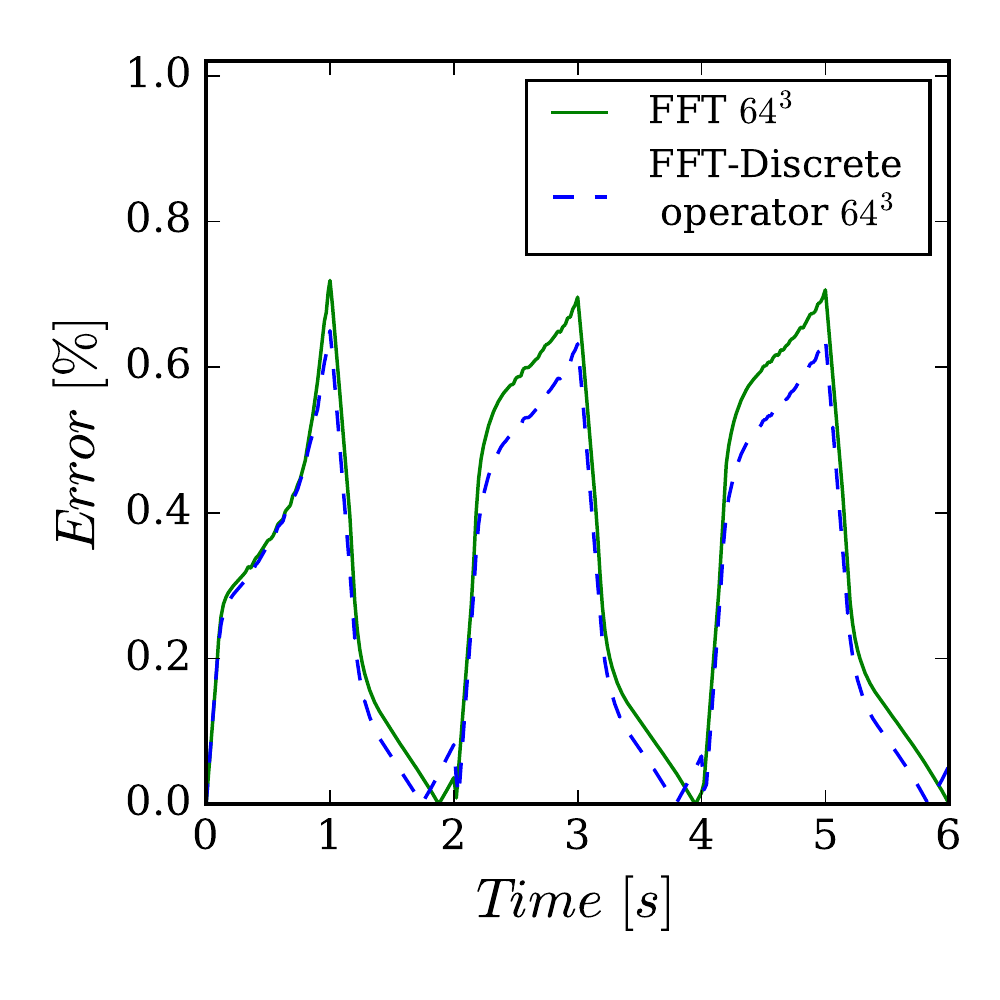}
\includegraphics[width=.32\textwidth]{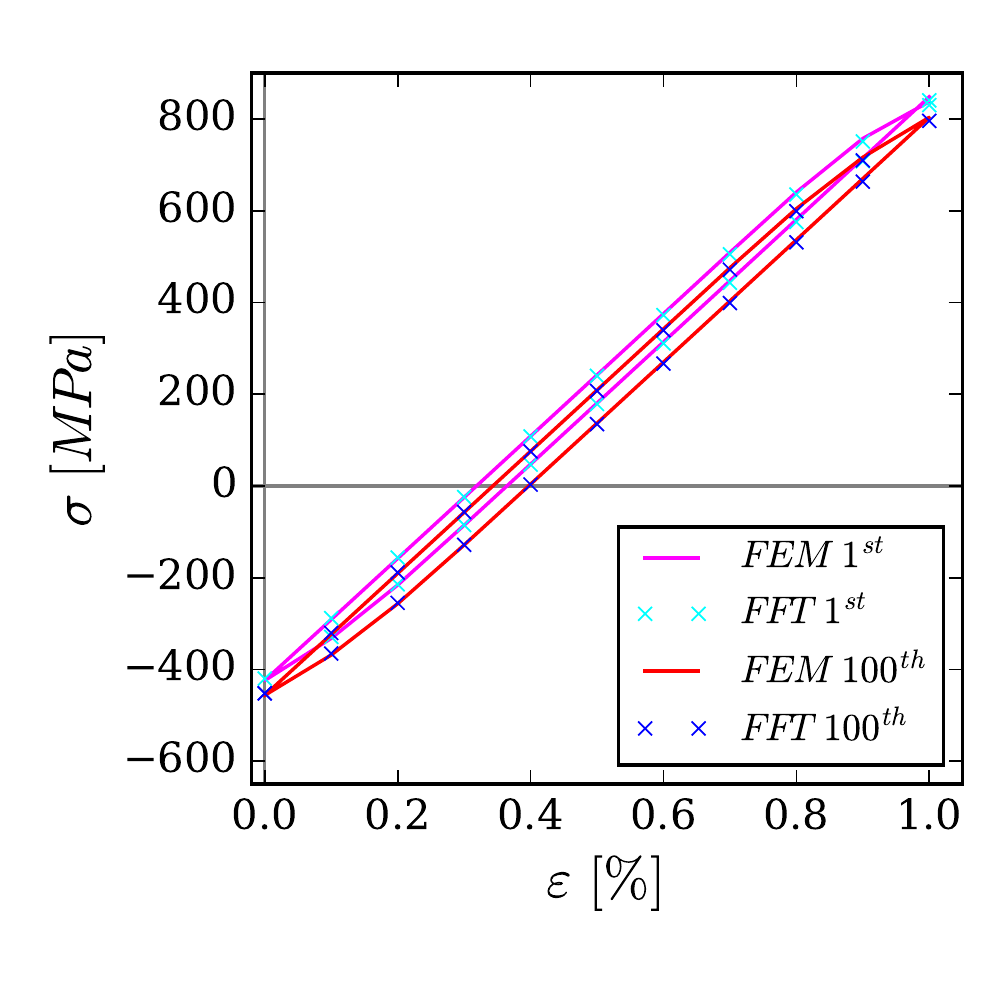}
\includegraphics[width=.32\textwidth]{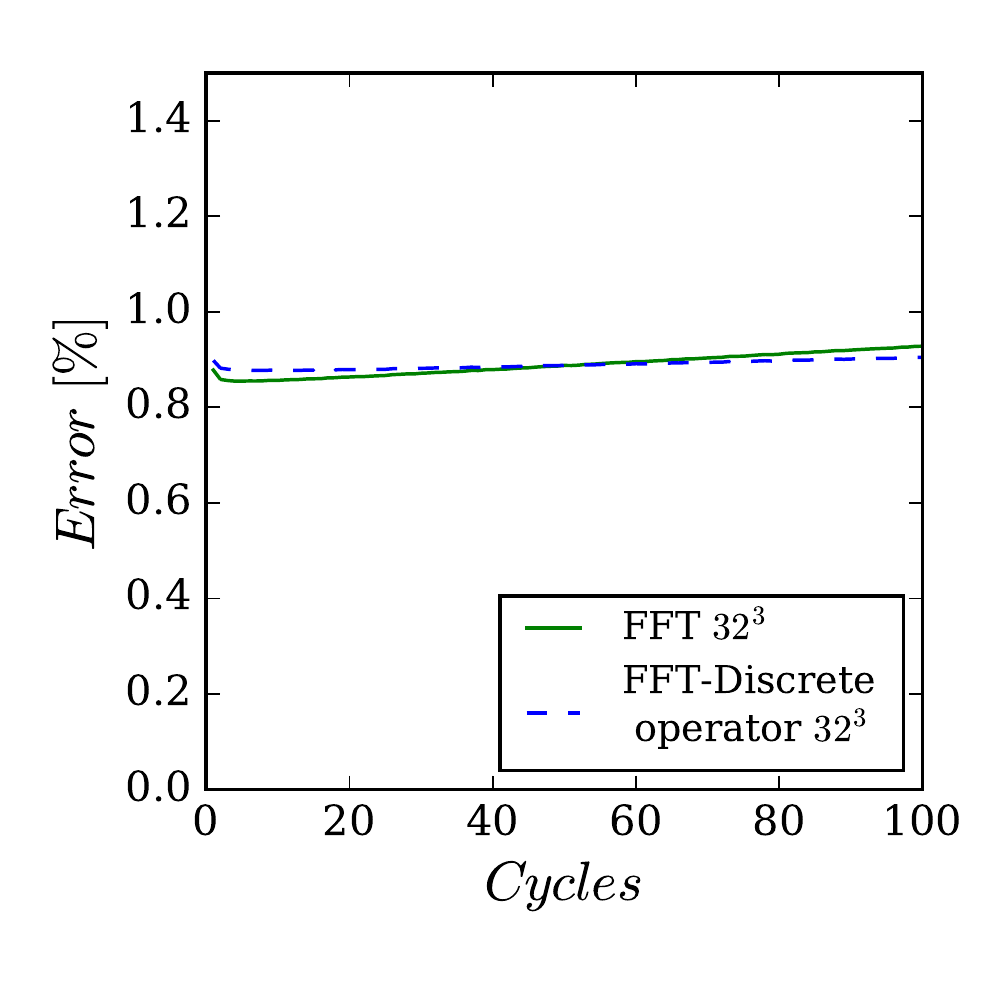}
\caption{(a) Error during 3 cycles of $\varepsilon_{max}\approx 5\%$ test in $64^3$  voxel/elements, (b) first and last cycle for a cyclic history of 100 cycles ($\varepsilon_{max})= 1\%$) in $32^3$ voxel/elements (c) evolution of error with the cycles corresponding to the previous load history}
\label{macro2}
\end{figure}

In the first case, the relative error in the macroscopic response respect the FEM result for the model with $64^3$ voxels  (eq. \ref{eq:error}) is represented in Fig. \ref{macro2} (a). It can be observed that the maximum difference is below 0.8$\%$ for the full load history being this difference smaller for lower stress levels. Moreover, it can also be observed that there is almost no error increase from cycle to cycle. In the second case a numerical experiment under strain amplitude of $\varepsilon_{max}=1\%$ for 100 cycles is performed using the small models with $32^3$ voxel/elements due to the computational cost of FEM simulations. In this simulation the stable cycle is reached and its shape is represented in Fig. \ref{macro2} (b) for both FEM and FFT. The stable cycle is perfectly captured by the FFT solver showing that there is no substantial error accumulation with the number of cycles. To highlight the stability of the solutions with the number of cycles,  Fig. \ref{macro2} (c) represents the difference between FEM and FFT solutions as function of the cycle number. It can be observed that this difference does not increase at all with the number of cycles when using the discrete projection operator. In the case of  the standard approach the error increased but very slowly, at a rate of 0.04\% for 100 cycles, also allowing to reach a huge number of cycles with a very small difference in behavior.

\subsubsection{Microscopic fields comparison}
\begin{figure}[H]
\includegraphics[width=.29\textwidth]{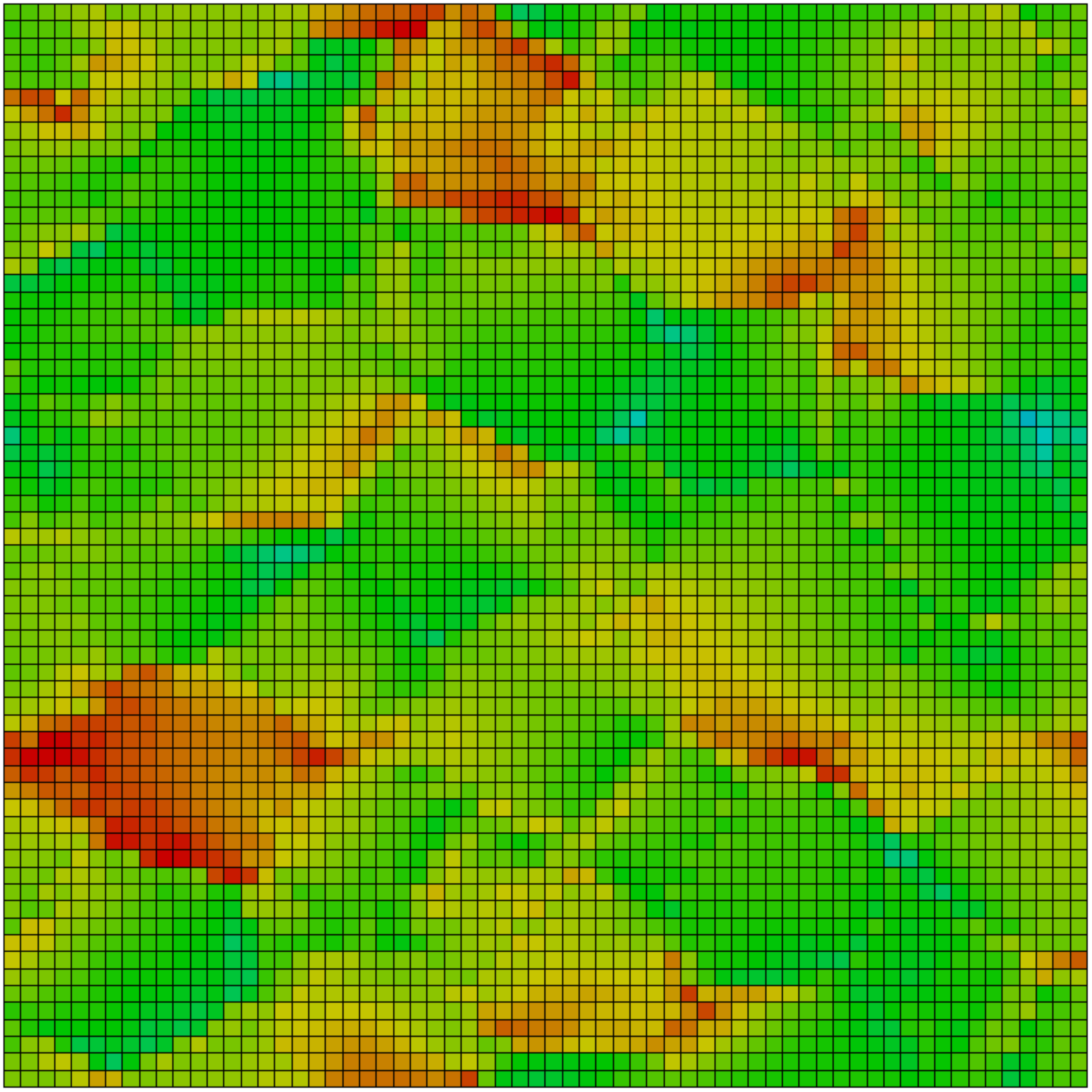}
\includegraphics[width=.29\textwidth]{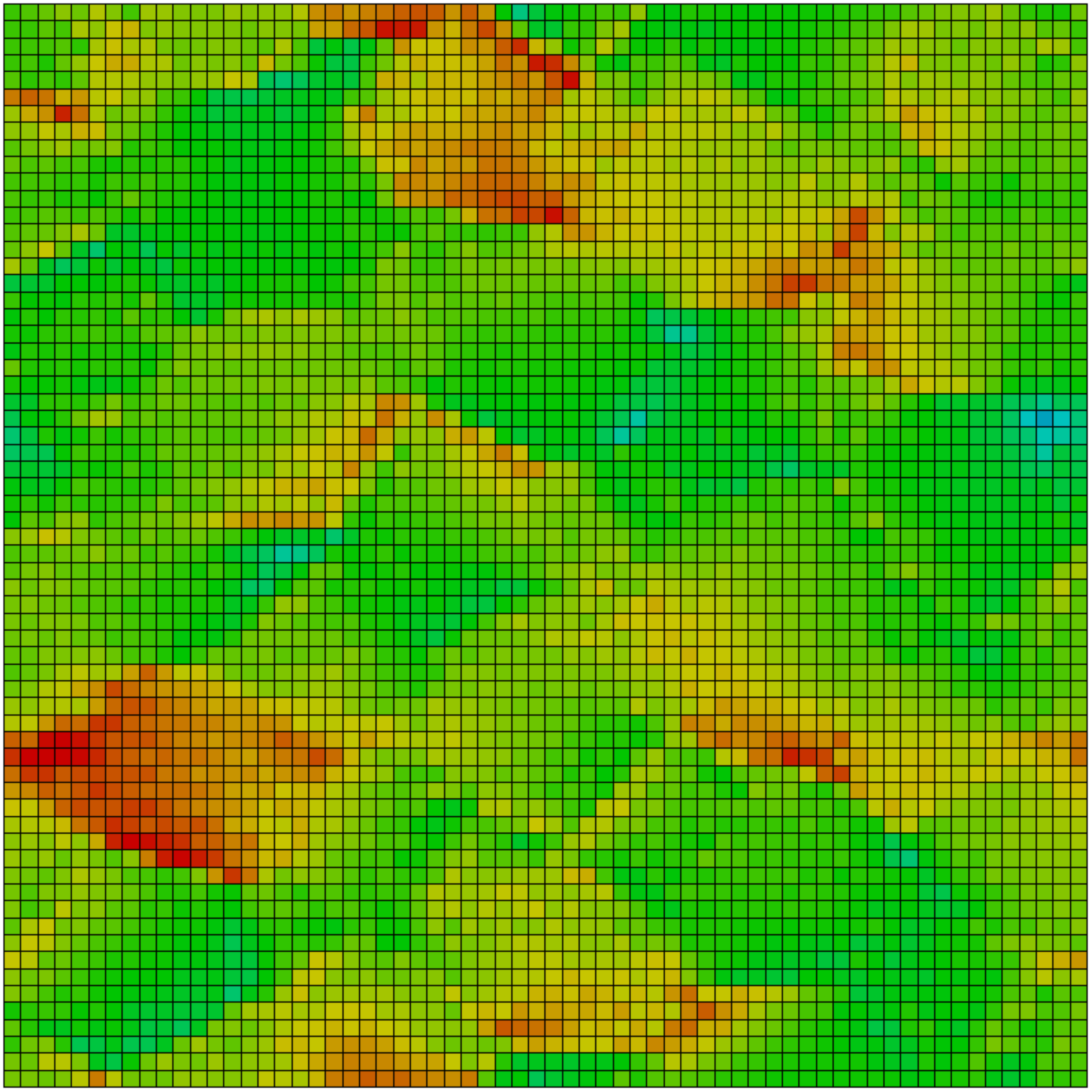}
\includegraphics[width=.29\textwidth]{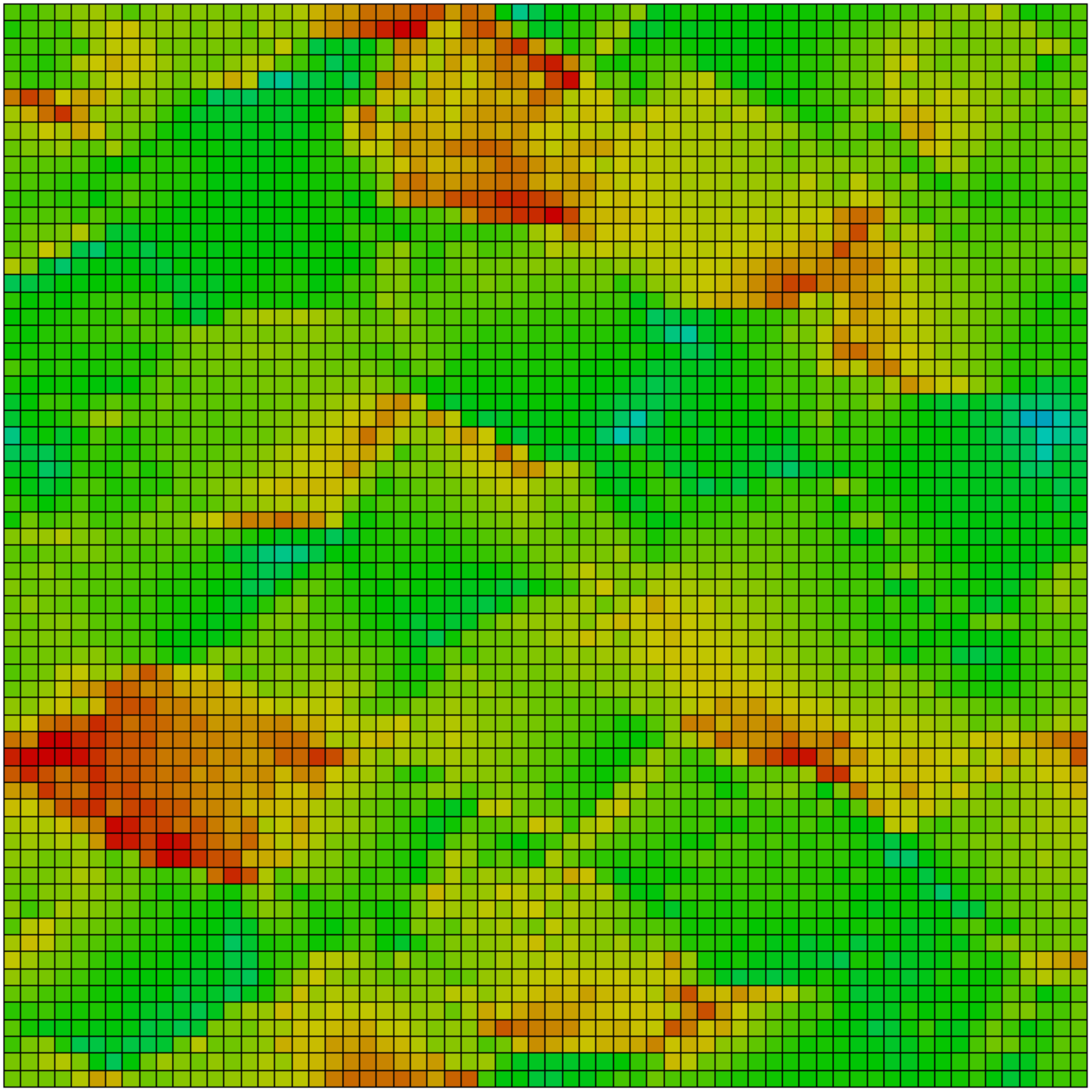}
\includegraphics[width=.09\textwidth]{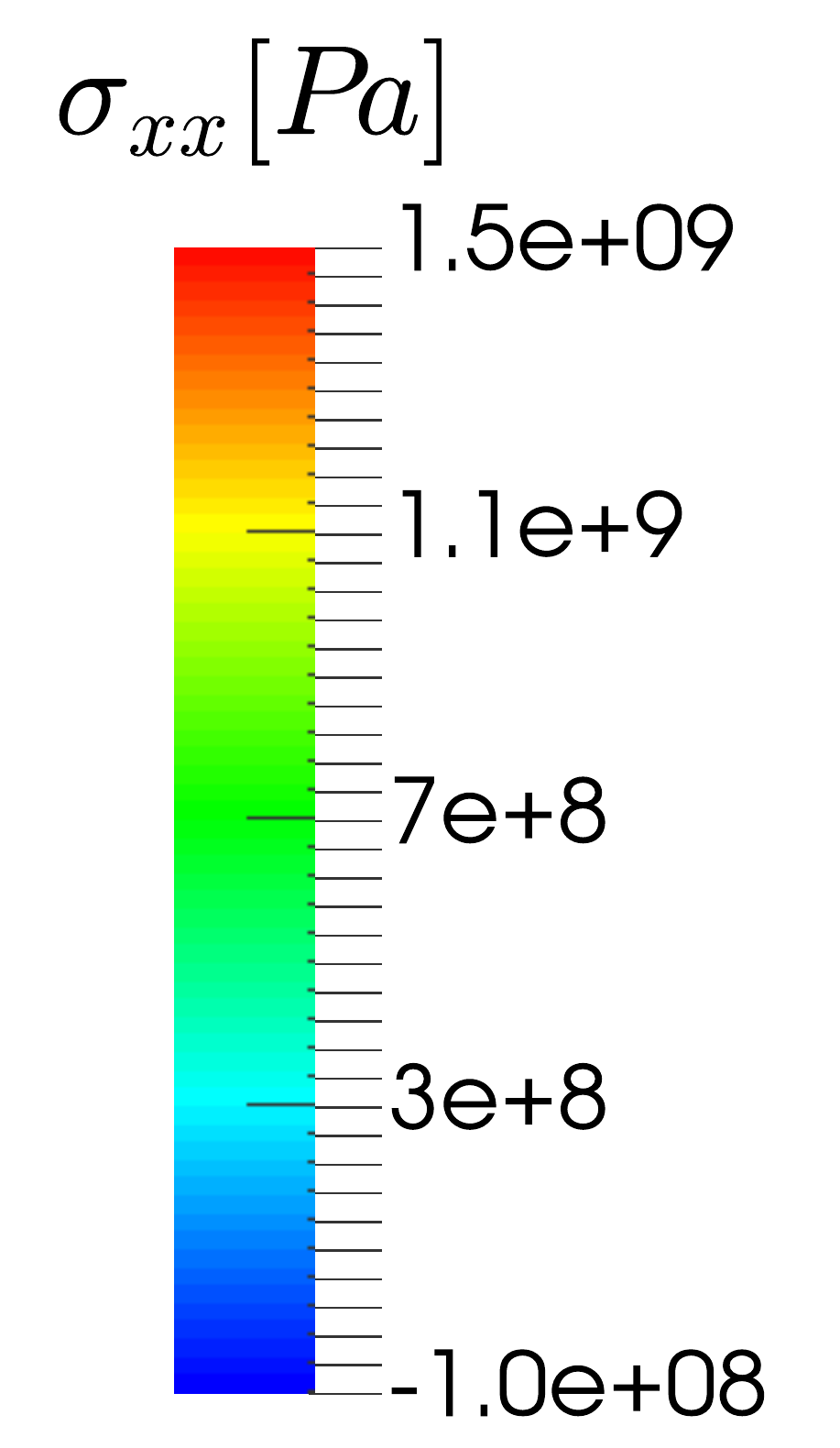}
\caption{Microscopic $\sigma_{xx}$ values in a RVE face, left: FEM results, center: FFT results, right: discrete operator version of FFT }
\label{fig:stress}
\end{figure}

The use of FFT for microstructure based fatigue modeling implies an accurate prediction of the microscopic fields because the fatigue life is estimated from the extreme values of those fields after a large number of cycles. To assess the ability of the FFT framework to predict accurate values of the microfields, the microscopic Cauchy stress in the loading direction, $\sigma_{xx}$, and the local value of the accumulated shear (fatigue indicator parameter $P_{cyc}$) are compared with the finite element solution. The comparison is presented for the models with $64^3$ voxels. The distribution of the stress in the loading direction after the first ramp-up load obtained with FEM, FFT and the discrete projection operator version of FFT is represented in Fig. \ref{fig:stress}. The figure shows the results in one of the RVE faces using the same code for the three simulations. The results are not smoothed out and correspond to the voxel value in the case of FFT and to the average element value in the case of FEM. 
The stress field obtained using the FFT approach is qualitatively very similar to the one obtained with FEM, as it can be observed in Fig. \ref{fig:stress}. Respect the two different projection operator used, the local stress maps are (again qualitatively) very similar and only some differences can be found in areas with stress concentrations.  

To quantify the differences between any local field $f(\mathbf{x})$ obtained with FFT with the FEM simulation result, $f_{FEM}$, the normalized $L2$  norm of the difference is proposed.
\begin{equation}
Diff(f)[\%]=\frac{||f -f_{FEM}||_{L2}}{||f_{FEM}||_{L2}}\cdot 100 \text{.}
\label{eq:L2}
\end{equation}
For the stress component parallel to the loading direction this difference in the $32^3$ models corresponded to 4.6\% and 3.6\% for the standard and discrete projection operator respectively. For the finer discretization of $64^3$ voxels these differences decreased to  3.4\% and 2.6\% respectively.  Clearly the use of the discrete projection operator in the FFT simulations provided local fields more similar to FEM solution. In addition, as expected, the differences between FFT and FEM are reduced with the mesh refinement.

\begin{figure}[H]
\includegraphics[width=.29\textwidth]{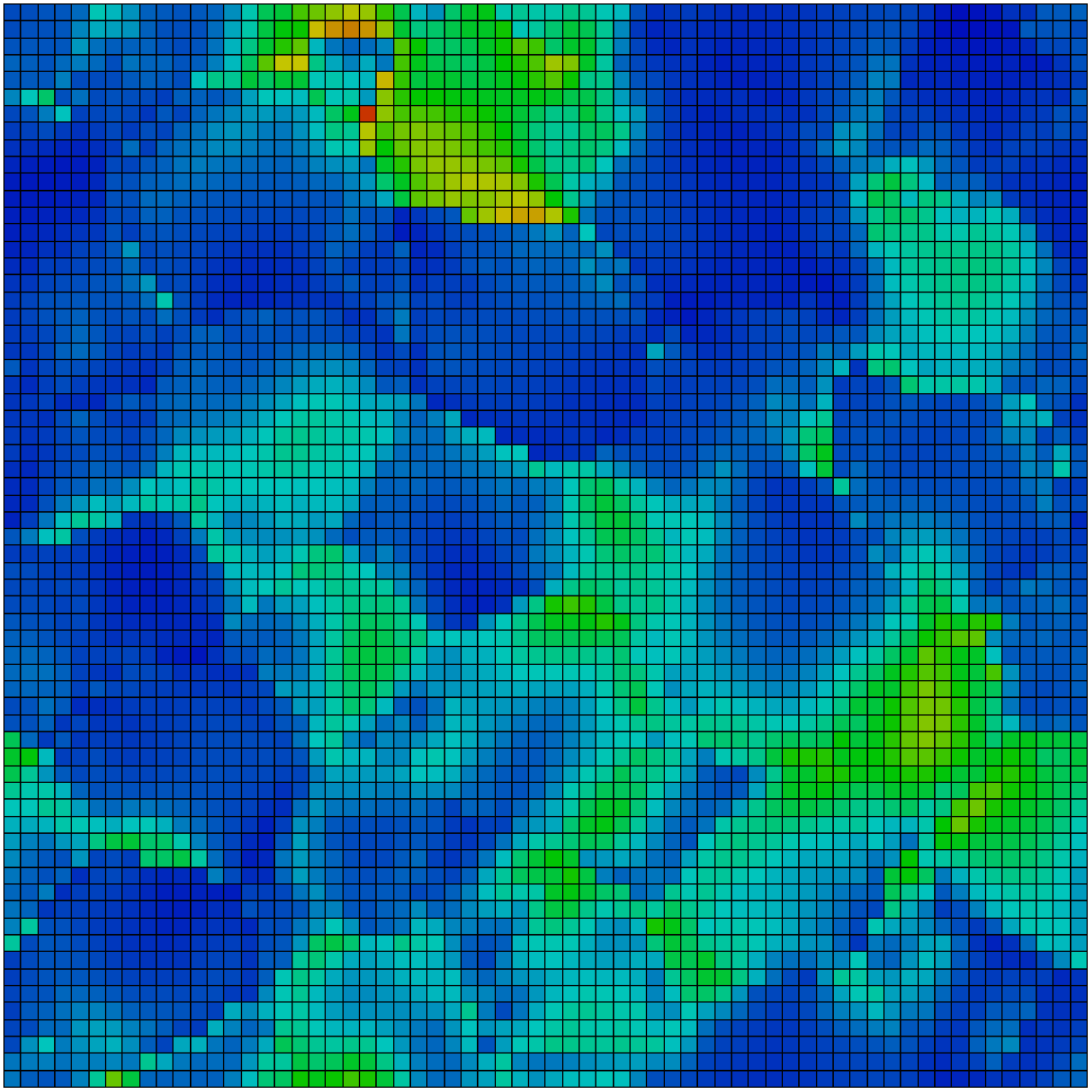}
\includegraphics[width=.29\textwidth]{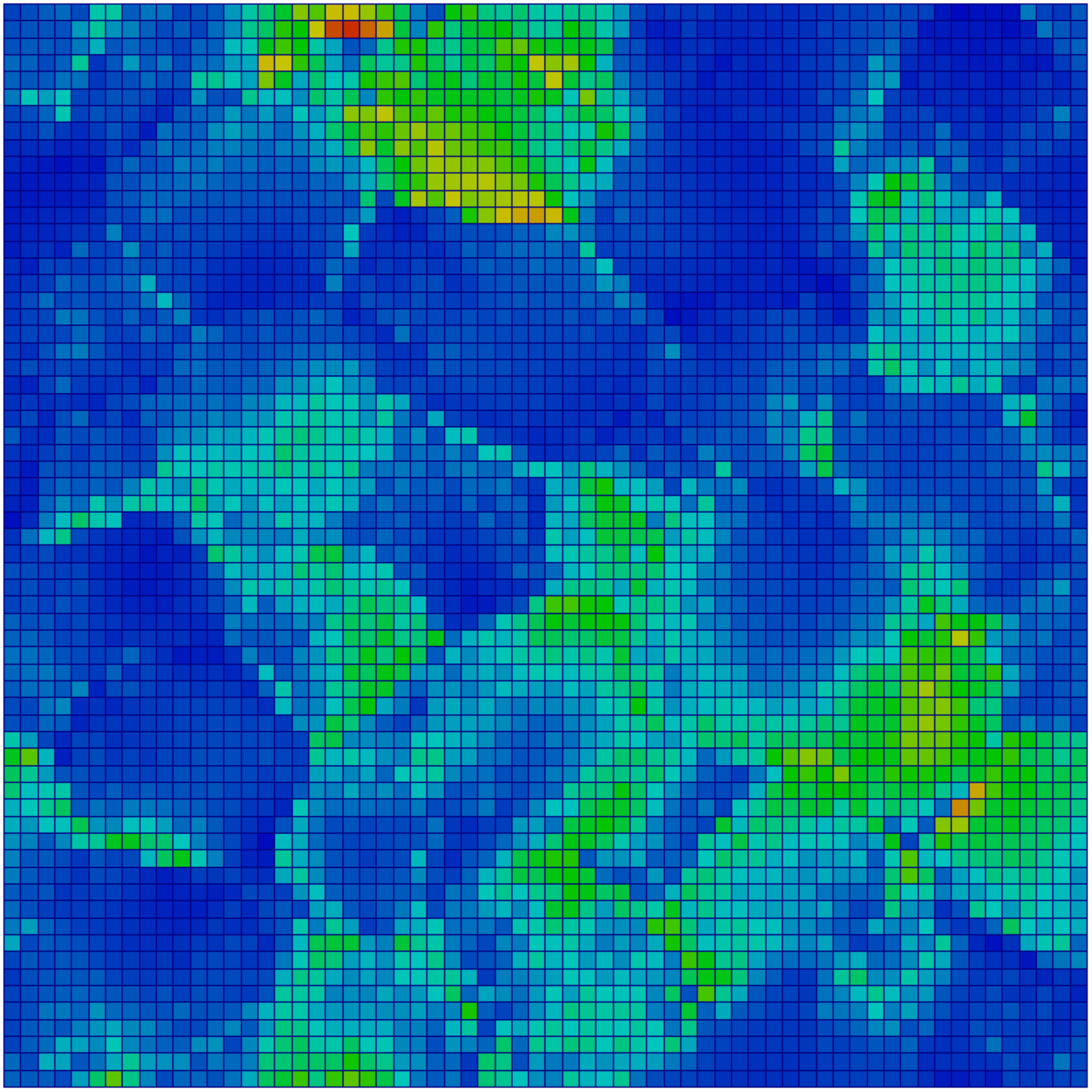}
\includegraphics[width=.29\textwidth]{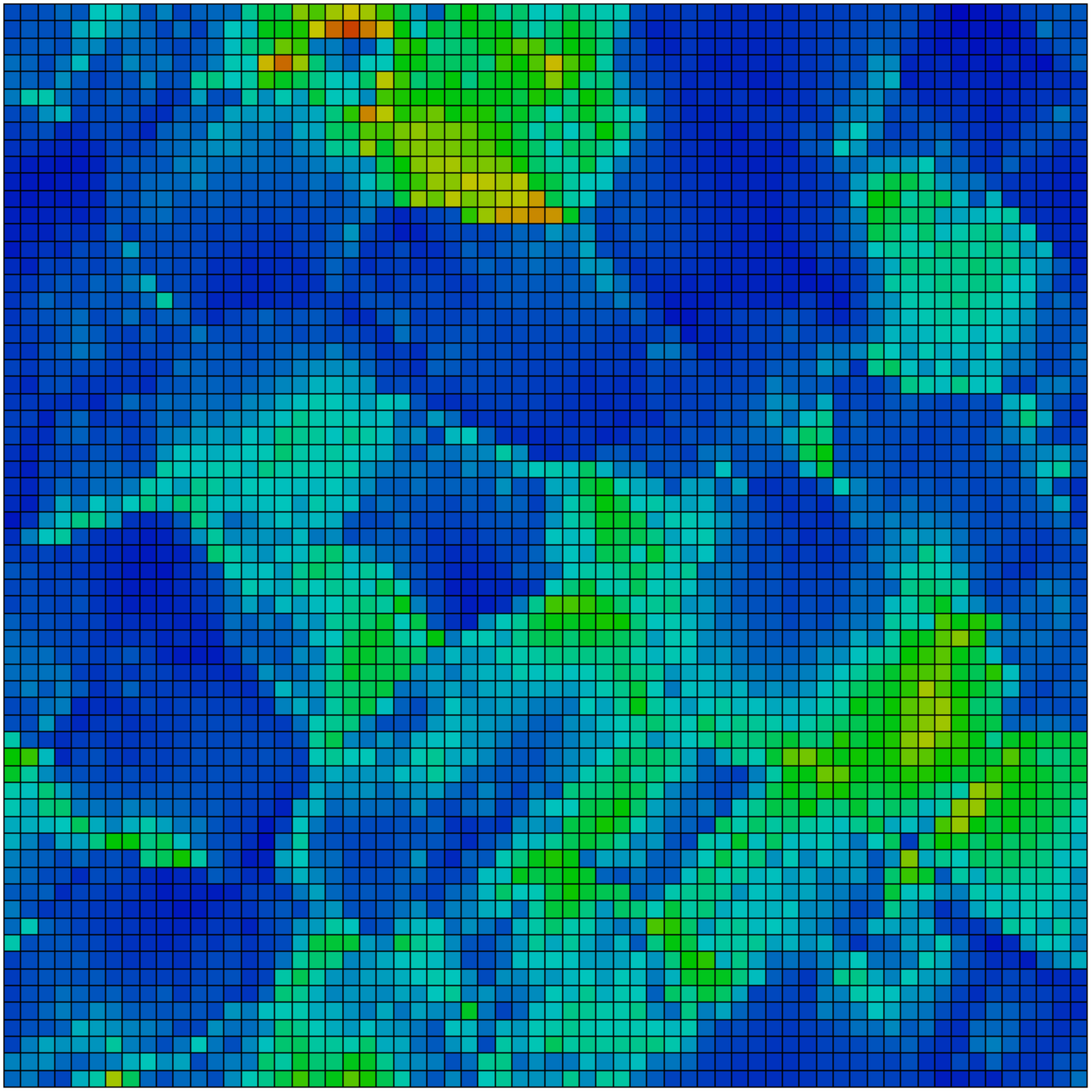}
\includegraphics[width=.09\textwidth]{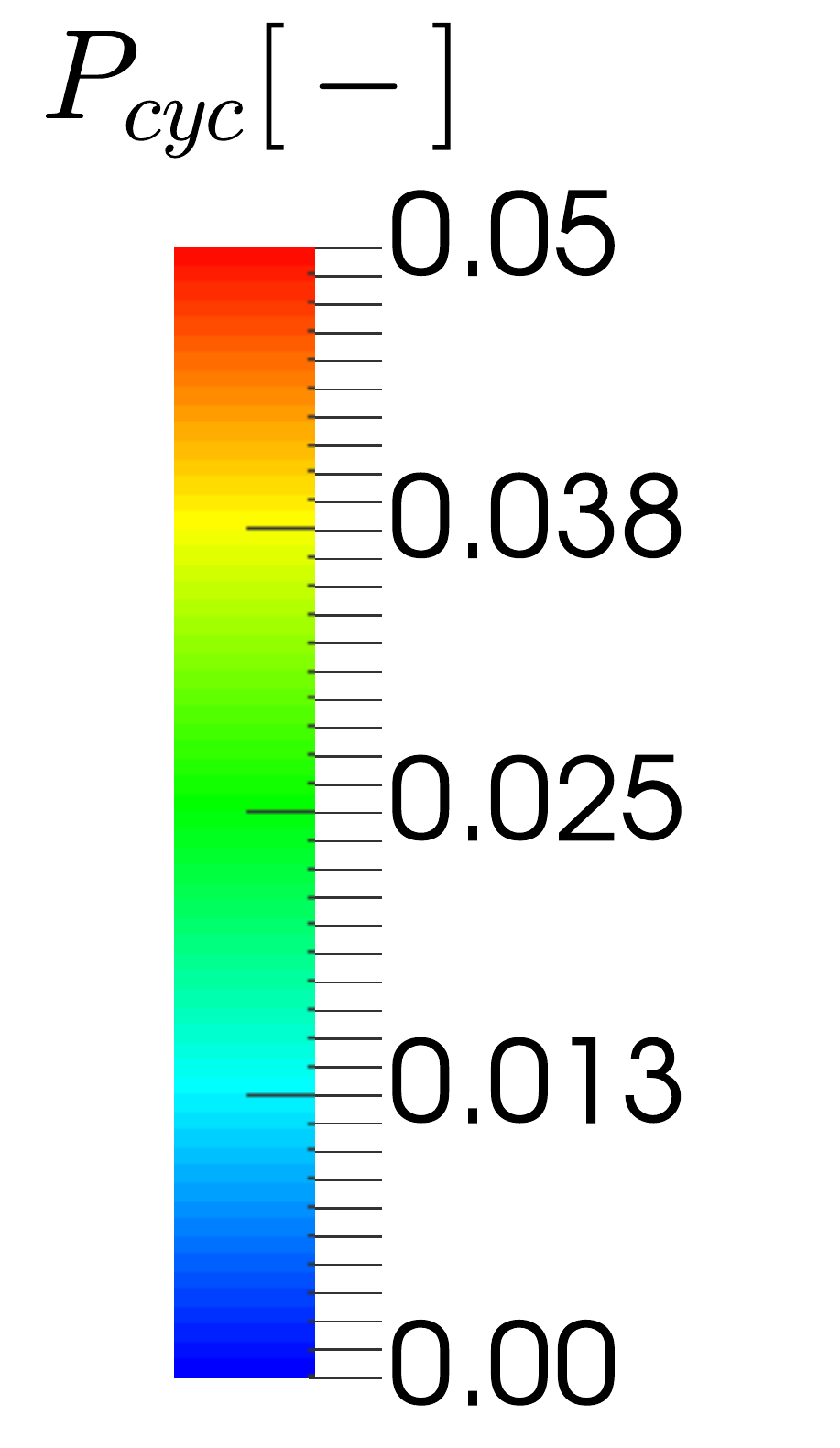}
\caption{Microscopic values of $P_{cyc}$ in a RVE face, left: FEM results, center: FFT results, right: discrete operator version of FFT}\label{fig:fip}
\end{figure}

The local value of $P_{cyc}$ in one of the RVE faces obtained after three cycles using the three different modeling approaches is represented in Fig. \ref{fig:fip}. Again, the results represented are the local voxel/element values without smoothing. As it happens with the stress field, qualitatively the FIP patterns obtained with the three methods are very close. Qualitatively, the FFT solution seems to present slightly higher FIP concentrations than the FEM prediction and the use of the modified projection operator led to a smoother FIP field that is also closer to the FEM results.  The differences obtained for the $P_{cyc}$ were  30.9\% and 23\% in $32^3$ models for the normal and discrete version respectively, and 25.1\% and 18\% in $64^3$ models. The results obtained for the other FIP considered in this study, $W_{cyc}$, are very similar and omitted here for the sake of brevity. Note that differences in the local relative differences found are affected by the very small value of the resulting plastic strain for a strain level of $\varepsilon_{max}=1\%$. Indeed, this error definition will become singular for small applied macroscopic strains and it is expected that the error will be reduced for larger strain amplitudes. Finally, it should be highlighted that, again the use of discrete projection operators lead to a result more similar to FEM.

\begin{figure}[H]
\includegraphics[width=.49\textwidth]{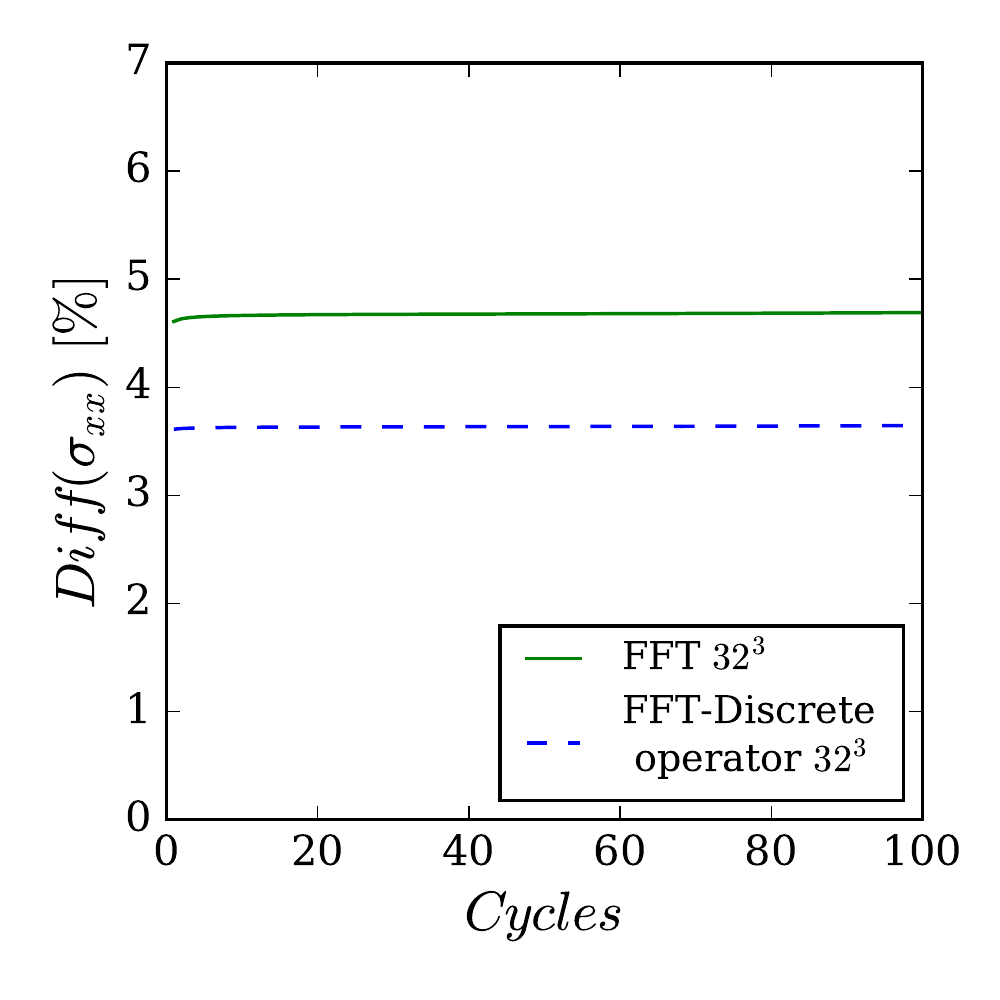}
\includegraphics[width=.49\textwidth]{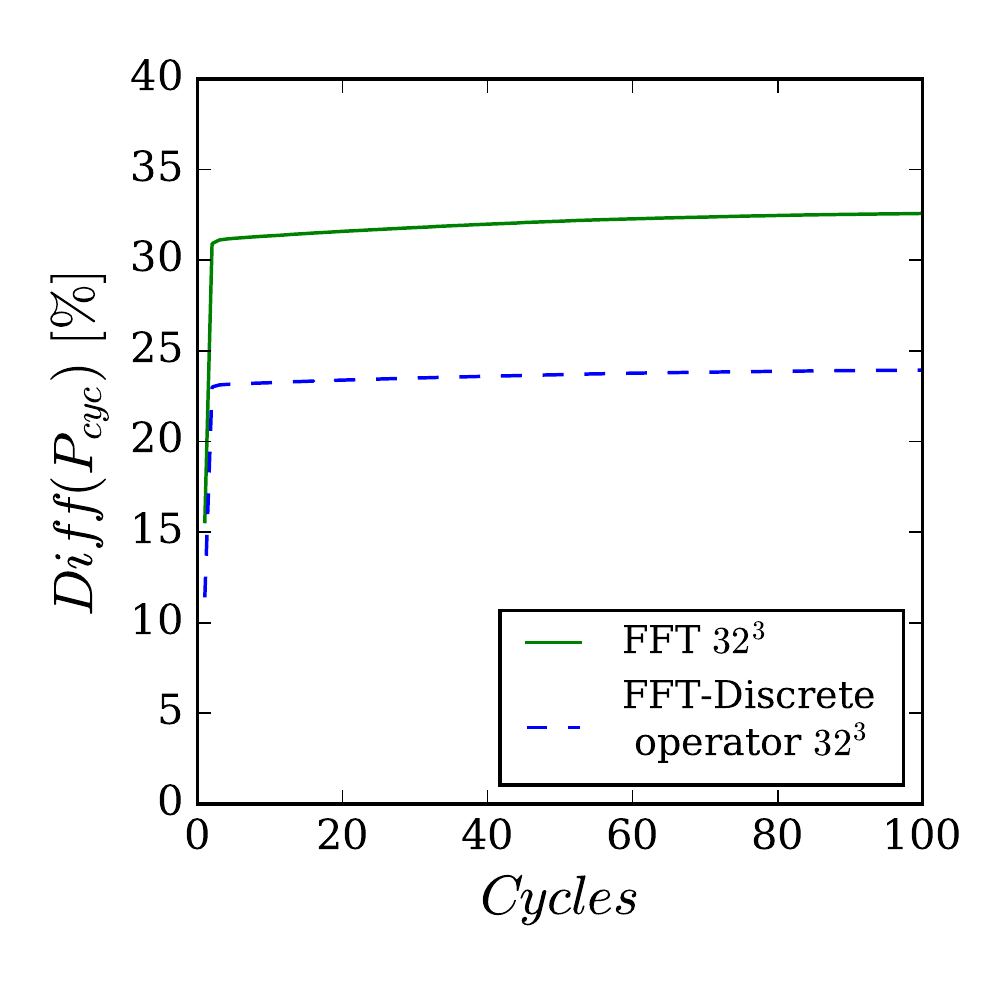}
\caption{(a) Difference of $\sigma_{xx}$ field during 100 cycles of $\varepsilon_{max}\approx 1\%$ test in $32^3$  voxel/elements and (b) Difference of $P_{cyc}$ field during 100 cycles of $\varepsilon_{max}\approx 1\%$ test in $32^3$  voxel/elements}
\label{diffevol}
\end{figure}

As it was done with the macroscopic response, the microscopic response is evaluated for two additional cyclic loading conditions, the application of a large number cycles and the increase in the applied strain, to asses the response of the FFT framework under different load histories. The evolution of the local field differences between FEM and FFT computed using (eq. \ref{eq:L2}) for 100 cycles and $\varepsilon_{max}=1\%$ is represented in Fig. \ref{diffevol} for the model with $32^3$ discretization. It can be observed that, as happened with the macroscopic response, both the differences in stress and strain maps are kept almost constant for all the cycles studied. The effect in the local fields of applying a larger value of the strain in each cycle was also studied using a strain range of $\varepsilon_{max}=5\%$ and the finer models ($64^3$ voxels). The  differences of the stress  between FFT and FEM obtained using eq. (\ref{eq:L2}) corresponds now 4.7\% and 3.3\% (standard and discrete projection operators), slightly greater than the ones obtained for $\varepsilon_{max}=1\%$. On the contrary, the relative differences in $P_{cyc}$ are notably reduced respect the ones obtained for  $\varepsilon_{max}=1\%$ being  now 5.3\% and 3.7\% for the standard and discrete operator respectively. This error reduction is due to the larger value of plastic strain produced for that level of applied strain.

\subsection{Fatigue life estimation}

The local values of the fatigue indicator parameters are not adequate to correlate the fatigue life of a given microstructure due discretisation errors. In order to reduce this effect, fatigue indicators are averaged in some regions to produce non-local measures, which are more adequate to estimate the tendency to nucleate a crack \cite{Castelluccio2014,Cruzado2017}. Therefore, it is expected that the small differences found between the FEM and FFT FIP microfields, after averaging, become much smaller. Moreover, the FIP representative of the full RVE and responsible of the fatigue life predictions is based on the tails of the non-local fatigue indicator parameter distribution so the differences in the microfields in regions with small FIP value will not affect the fatigue life estimation. 
In this section the differences between the resulting FEM and FFT non-local fatigue indicators parameter distribution will be studied and the implication of this difference in the RVE fatigue life estimation will be analyzed. 

\begin{figure}[H]
\includegraphics[width=.49\textwidth]{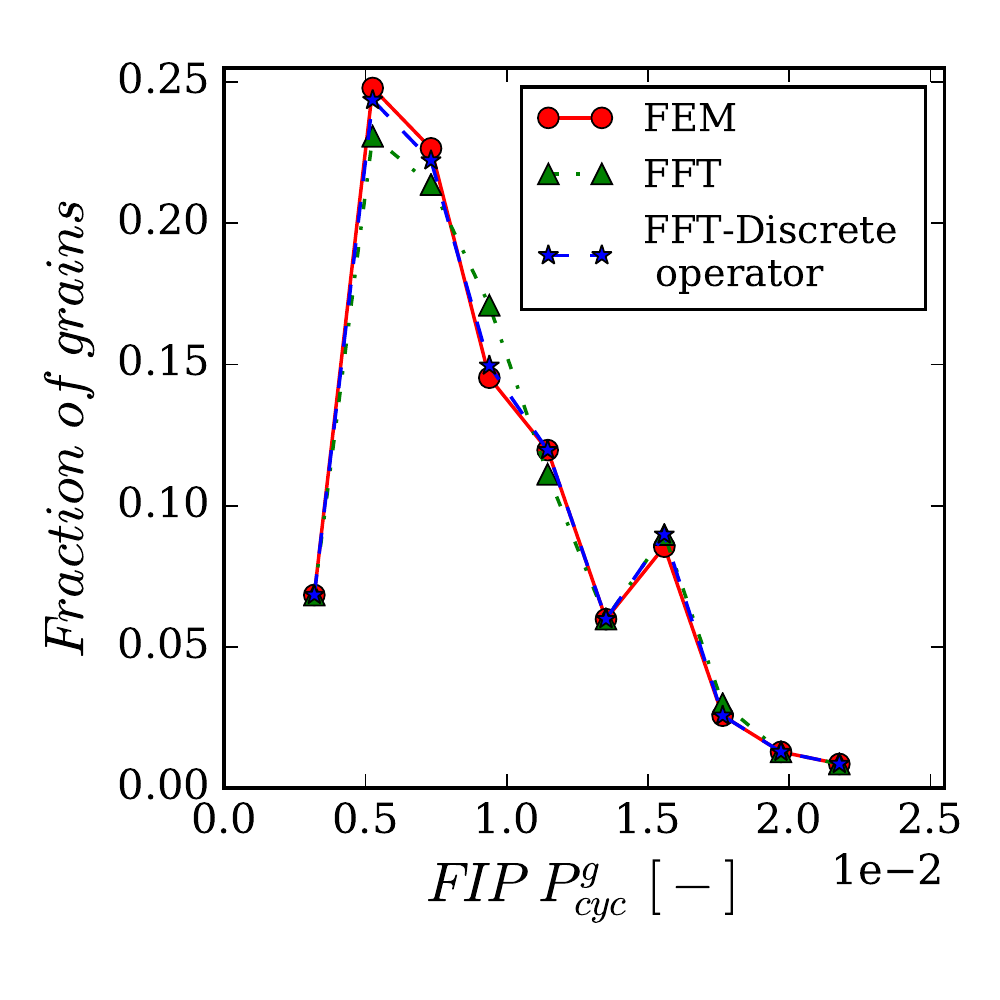}
\includegraphics[width=.49\textwidth]{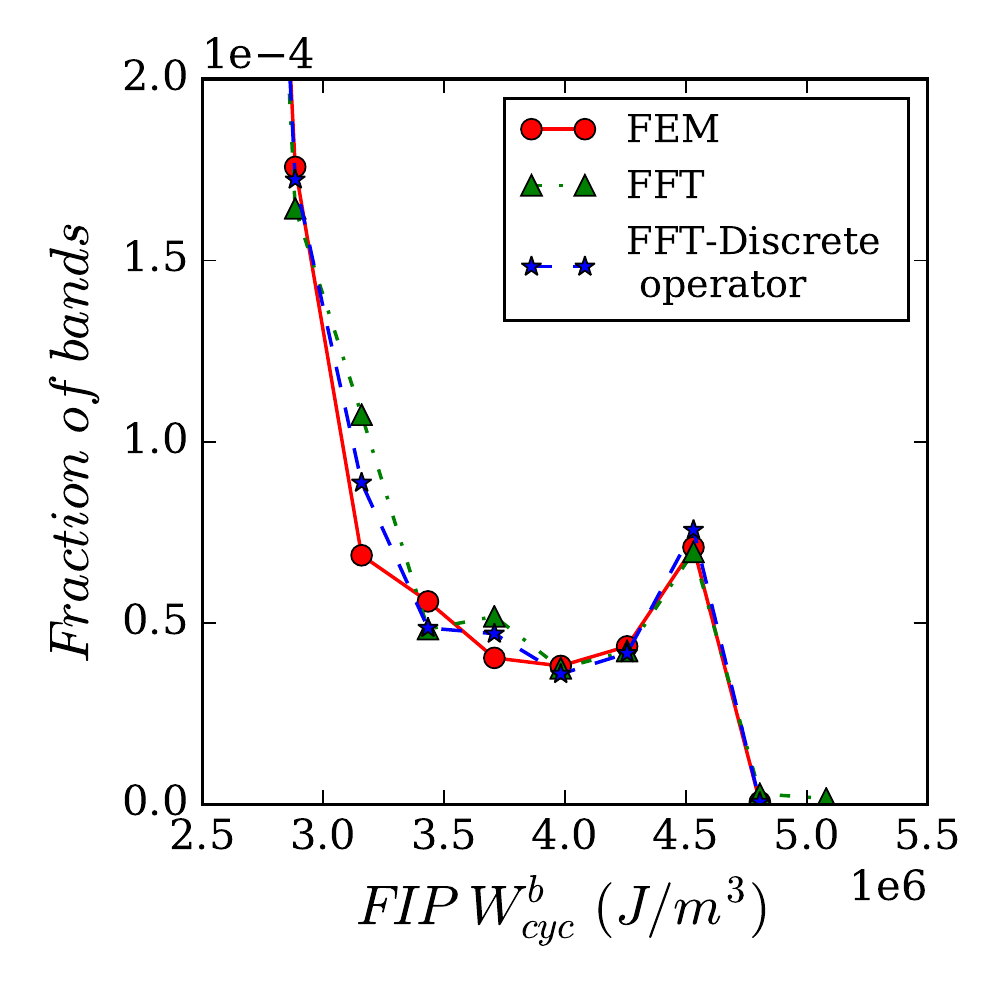}
\caption{FIP histogram for $64^3$ voxel model in FEM, FFT, and discrete projection operator FFT. (a) grain averaged $P_{cyc}$ and (b) band averaged $W_{cyc}$}
\label{fig:histo1}
\end{figure}

Two non-local fatigue indicator parameters will be considered, grain averaged $P_{cyc}$ and the band averaged $W_{cyc}$ and simulations will be done with the $64^3$ voxel model. In the case of the grain averaged $P_{cyc}$, one FIP value is obtained per grain (here 235 values) and a histogram showing the fraction of grains as function of the $P_{cyc}$ is represented in Figure \ref{fig:histo1}(a). A similar graph is computed for $W_{cyc}$ and represented in Fig. \ref{fig:histo1}(b), but in this case the histogram represents the fraction of averaging bands.
When the FIPs are averaged at grains (Figure \ref{fig:histo1}(a)) it can be observed that the distribution of the local FIP through the grains obtained using FEM and FFT are very similar, presenting only some small differences for small FIP values (left part of the curve). It can also be observed that when FFT is based on the discrete projection operators the curves become almost identical in the full FIP range. This result highlights the conclusion obtained when comparing the local fields point by point: the modified projection operator produce a response in all the microfields studied (stress, strain and the two FIPs considered) more similar to FEM than the standard FFT method. The histogram with the volumetric distribution of $W_{cyc}$ (\ref{fig:histo1}(b)) shows more differences between FEM and FFT results and, again, the FFT fields obtained with the modified projection operator are closer to the FEM solution. This better agreement with FEM obtained using the modified projection operator is very clear in the tail of the distribution for large FIPs, where standard FFT predicts a maximum FIP considerably larger than the other two methods.
 
Next, the FIP representative of the microstructure ---in which fatigue life prediction is based on--- is obtained using FE, FFT and FFT based on the discrete operator. To compute this value, ten different RVEs representative of the same microstructure and containing $64^3$ voxels are simulated (the full set is some times called a Statistical Volume Element or SVE \cite{Castelluccio2014}). Each RVE followed the same grain size distribution previously described and has around 230 grains. The simulations consist in three cycles following the strain history defined in equations (\ref{eqFt}) and (\ref{eqFt2}). The maximum FIP of each RVE is extracted from the simulation using equations (\ref{Pmax}) and (\ref{Wmax}) and the FIP representative of the microstructure is obtained as the average of the FIP of each RVE (ten in total). The relative difference between the two FFT approaches and the reference FEM solution is obtained as
\begin{equation}
Err(\%)=100\frac{FIP_{FE}-FIP_{FFT}}{FIP_{FE}}
\label{eq:dif}
\end{equation}
where $FIP_{FE}$ and $FIP_{FFT}$ stand for the averaged FIP, representative of the full microstructure. The average FIP value, the standard deviation and the relative difference respect FEM  are represented in Table \ref{table:errorsFIP1} for grain averaged $P_{cyc}$ and in Table \ref{table:errorsFIP2} for band averaged $W_{cyc}$.
\begin{table}[H]
\centering
\begin{tabular}{|c|c|c|c|}
\cline{1-4}
 & FEM  & FFT &  FFT discrete operator \\ \hline
mean []  & 2.567e-2   & 2.603e-2  &   2.597e-2 \\
\hline
std deviation []  &  6.021e-3   &  5.845e-3 &    5.942e-3 \\
\hline
difference with FEM (eq. \ref{eq:dif}) & 0 & 1.40\% & 1.17\% \\
\hline
\end{tabular}
\caption{Mean value, standard deviation and difference with FEM of the microstructure representative value of grain averaged $P_{cyc}$}
\label{table:errorsFIP1}
\end{table}

\begin{table}[H]
\centering
\begin{tabular}{|c|c|c|c|}
\cline{1-4}
 & FEM  & FFT & FFT discrete operator \\ \hline
mean [N/m]  & 5.071e6  &  5.823e6 &   5.424e6 \\ \hline
std deviation [N/m]  &  5.112e6   & 8.763e5    & 6.948e5\\ \hline
difference with FEM (eq. \ref{eq:dif})  & 0 & 14.6\% & 6.8\% \\
\hline
\end{tabular}
\caption{Mean value, standard deviation and difference with FEM of the microstructure representative value of grain averaged $W_{cyc}$}
\label{table:errorsFIP2}
\end{table}
The results on Table \ref{table:errorsFIP1} shows that, when grain averaged is used, the value of $P_{cyc}$ representative of the microstructure and obtained using either FEM and FFT are almost identical (differences below 2\%), being the value of FFT method slightly larger than FEM in all the cases. Respect the standard deviation of the ten FIP values obtained, FEM and FFT results are very similar being the scatter found in FFT slightly smaller. In this case the use of a modified projection operator  has a small influence in the results. The reason for the strong similarity between FEM and FFT is that grain averaging implies integration over many material points and therefore the local differences found are smooth out in the averaging procedure. The second FIP studied, $W_{cyc}$, is obtained by band averaging  and the resulting FIPs and comparison with FEM are shown in Table \ref{table:errorsFIP1}.  It can be observed that the differences are higher compared to the grain averaged $P_{cyc}$, but they are still quite limited, being always below 15\%. Again, FFT results are slightly larger than FEM solution. The dispersion between the ten RVEs is larger than for grain averaging, an expected result because the non-local smoothing procedure is done here in bands containing much less element than a grain. Nevertheless, the most interesting result here is that the introduction of the modified projection operator in the FFT simulation significantly reduces the error with the reference FEM solution to less than the half, from 14.6\% to 6.8\%. 


Finally, the influence of the differences found in the FIP representative of the microstructure between FEM and FFT will be used to estimate the deviations expected in the fatigue life (number of cycles to nucleate a  crack) using both techniques. In light of the previous results it is clear that the modified version of the projection operator have to be used to minimize the difference in the FIP values representative of the microstructure with respect FE. If the fatigue life prediction law is adjusted using FEM models, this small differences in FIP will be translated to the fatigue life prediction. In the case of a linear relation to correlate FIP with life \cite{Manonukul2004,Sweeney2013}, the relative difference found in the FIP will be directly the difference in the number cycles for crack nucleation between FEM and FFT and will be limited in the worst situation to 6.8\% (band averaging). However, linear relations are not ideal to correlate fatigue because and more complex expressions are proposed. In particular, a power law with two parameters is used by other authors \citep{MCDOWELL20101521,Cruzado2018}  to correlate the number cycles to nucleate a crack $N$ with the $FIP_{cyc}$ 
\begin{equation}
FIP_{cyc}^m\cdot N = FIP_{crit}
\end{equation}
where the $m$ and $FIP_{crit}$ are material parameters, being $m$ an exponent with values between 1 and 2. Considering the maximum error in the $FIP_{cyc}$ prediction with respect FE, 6.8\%, the maximum relative deviation expected in fatigue life prediction is $\approx$15\%. This deviation will be well accepted for fatigue life prediction because the scatter in life of different fatigue experiments is far beyond this difference.

\subsection{Effect of second phases}
 The presence of stiff second phases as carbides in the alloy microstructure has a strong effect in the fatigue response due to the stress concentrations produced around them. Considering this situation in microstructure based fatigue life prediction is not a problem using FEM but FFT simulations using the standard projection operator might lead to Gibbs oscillations due to the strong phase contrast, and therefore local fields and FIPs might be inaccurate. The introduction of the modified projection operator proposed in this study and based on the discrete derivatives introduced in \cite{Willot2015} will allow to alleviate this problem and extend the use of FFT in fatigue to RVEs containing second phases. In order to assess the validity of this framework in the presence of hard phases, a distribution of particles ten times stiffer than the crystals are included in the polycrystalline model, occupying 1\% of the total volume. The particles are introduced as spheres with a diameter following a log-normal distribution of sizes of mean $4\mu m$ and $\sigma=0.56$ obtained from a experimental measure of carbides inside the microstructure of Inconel 718. In this case, models containg $64^3$ voxels are used and simulations are performed using FEM and FFT with the two projection operators considered.

\begin{figure}[H]
\includegraphics[width=.49\textwidth]{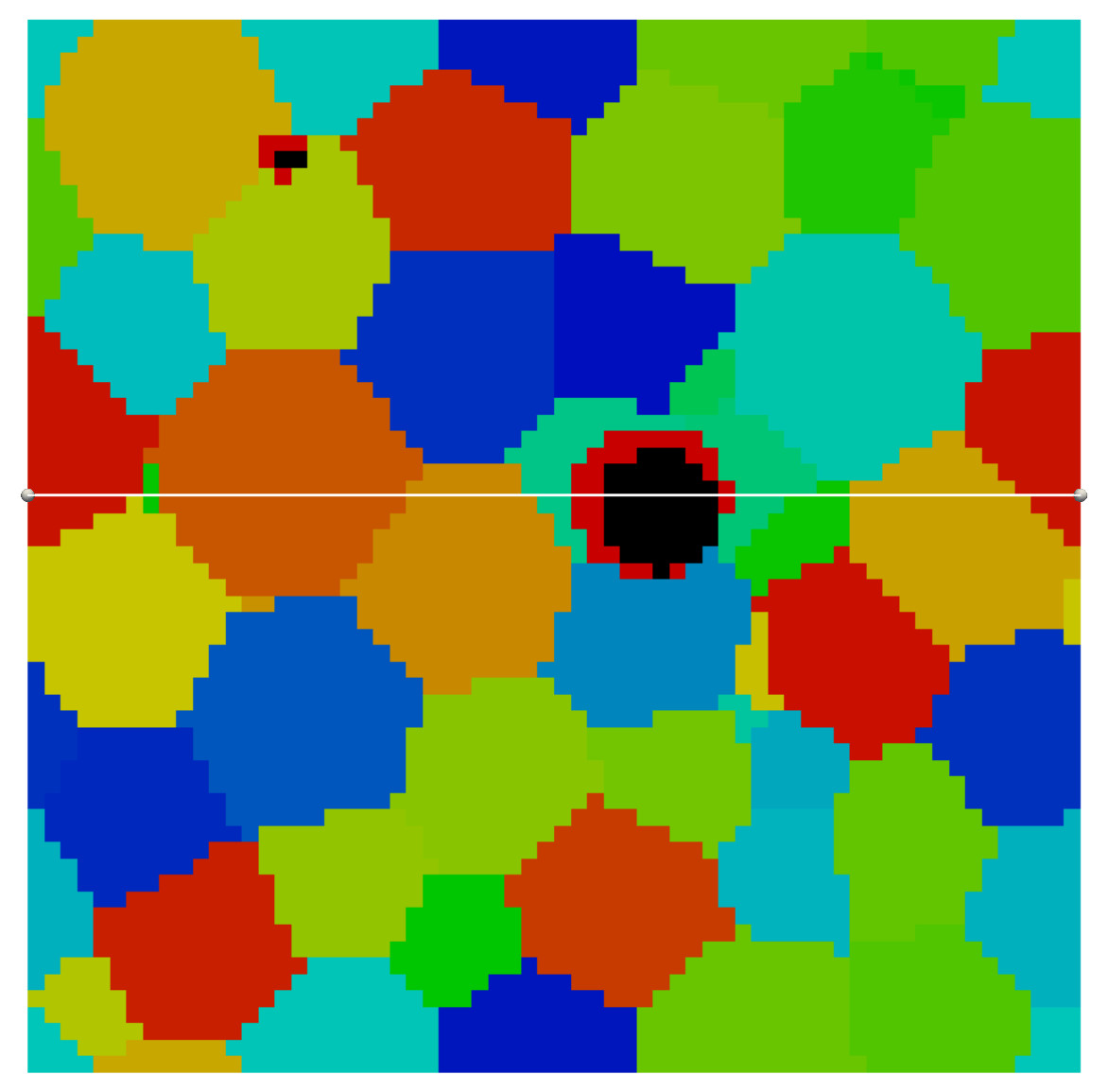}
\includegraphics[width=.49\textwidth]{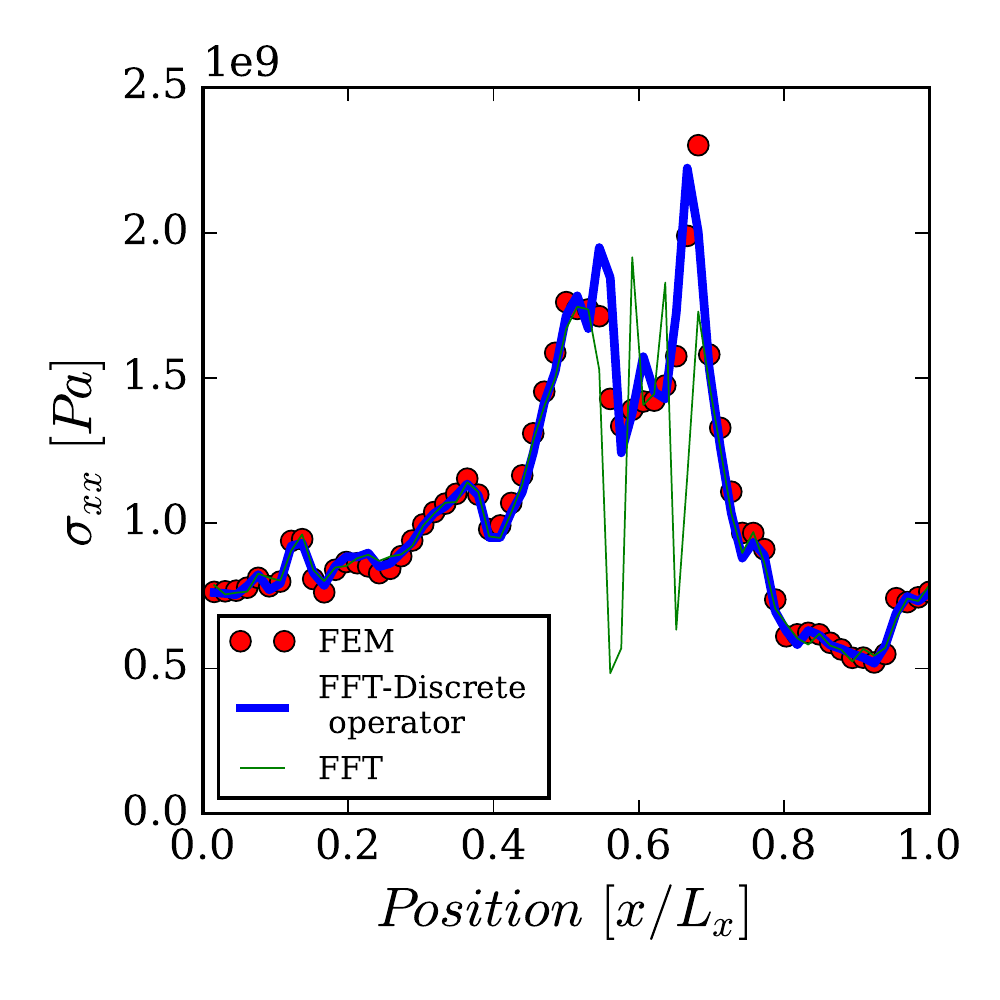}
\caption{(a) Microstructure with stiff particles (in black) and (b) $\sigma_{xx}$ along the line path}
\label{carb}
\end{figure}

An overview of the resulting stress field of the simulations with carbides is represented in Fig. \ref{carb}. A straight path crossing a stiff particle is shown as a white line (Fig. \ref{carb}(a)) and the value Cauchy stress component parallel to the applied deformation, $\sigma_{xx}$ along that path obtained using the three simulations is represented in Fig. \ref{carb}(b). In this figure the artificial oscillations in the response obtained using standard FFT are clearly shown and the strong improvement using the discrete projection operator can be appreciated.

From a quantitative view point, the presence of hard particles induce stress concentration around them and the average differences respect FEM obtained using the $L2$ norm slightly increased, resulting in 8.6\% and 6.9\% for the standard and discrete operator respectively. The difference between the plastic strain fields (represented by  $P_{cyc}$) obtained using FEM and FFT increased respect the ones found for pure polycrystals, from 25.1 to 29.8\% for the standard projection operator and from 18 to 22.8\% in discrete version. As in all the previous cases, the use of the discrete operator increased the similarity with FEM results. It must be finally noted that the local field difference is still considered small for fatigue life prediction, so the effect of considering second phases with one order of magnitude of stiffness contrast can be accounted accurately using this FFT approach.

\section{Analysis of computational performance}
The main reason for using FFT is the reduction in the computational cost respect to finite element simulations. It is well known that standard FFT approaches are much more efficient solving large linear models with limited phase contrast due to the order $n\log n$ growth of the computational cost. Therefore, elastic simulations or non-linear calculations for small applied strain values can be orders of magnitude faster when using FFT \citep{Eisenlohr2013}.   However, this improvement might be more limited in the case of micromechanics based fatigue simulations for many reasons, (1) the non-linear constitutive equation (cyclic crystal plasticity) is complex and very expensive to evaluate, (2) the simulations requiere many cycles including several path changes. The variational FFT framework used here might be an ideal framework to overcome these limitations thanks to absence of a reference medium, the better performance of the discrete operator and the use of  a fully implicit scheme using Newton-Raphson with consistent tangent matrices that allows reducing drastically the number of load increments. 

\begin{figure}[H]
\includegraphics[width=.8\textwidth]{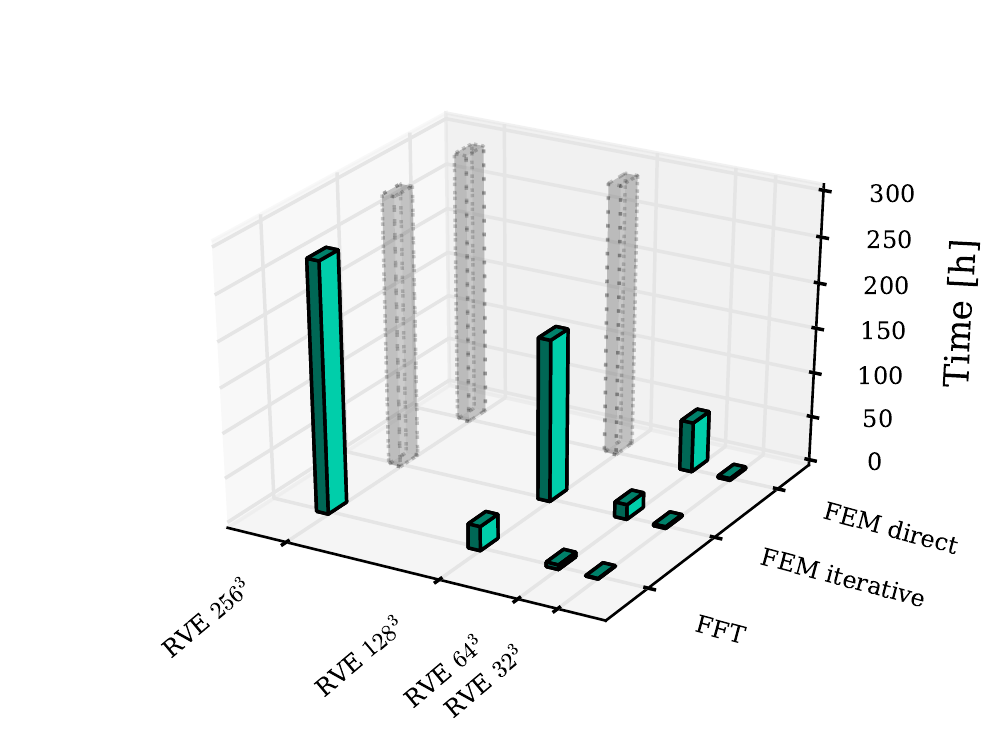}
\caption{Simulation time (in hours) for FEM and FFT simulations for different RVE sizes}
\label{perfo}
\end{figure}

 The comparison of the performance between FEM and FFT is done using one of the cases presented in the result section, the simulation of the alloy response for three full cycles under quasi-uniaxial loading, strain amplitude of $\varepsilon_{max}=1\%$ and $R_{\epsilon}=0$. Four different model sizes are considered $32^3$, $64^3$, $128^3$ and $256^3$. The simulations were performed in a 120GB RAM 2$\times$8 core Intel Xeon E5-2630 v3 workstation, using 4 cpus for this analysis. The FEM simulations were run using Abaqus/Standard 6.13.3 and the FFT simulation in our home-made code FFTMAD. In the case of FEM both the standard direct \emph{lu} solver and a Krylov iterative solver are used for the linear problem. Note that this last algorithm is very similar to the linear solver used in the  FFT framework so in this case both methods can be fairly compared. The total time needed for the simulation of each RVE size using both FEM and FFT is graphically represented in Fig. \ref{perfo}.

It can be observed that simulation times are similar for $32^3$ voxels while the FFT approach is 4 times faster in the $64^3$ model compared to the FEM iterative solver and around 10 times faster than FEM using a direct solver. In the case of $128^3$ voxel models the comparison with FEM was only possible with the iterative solver because of the memory requirement for the direct solver and in the largest model with $256^3$ elements the FEM code was not able to run the simulation with any type of solver. For the models with $128^3$ voxels (with more than 2 million voxels) FFT was 6 times faster and required approximately 4 times less memory and  than the FEM iterative simulations. These results show that this FFT framework is a reliable option to replace FEM for microstructure based fatigue simulation reducing the computational time and allowing to increase the size and the complexity of the RVEs to improve the accuracy in the fatigue life predictions.

\section{Conclusions}
A new framework is proposed specially adapted for micromechanics based fatigue simulation of polycrystalline metals using a spectral solver. This framework is based on the variational FFT approach \cite{Geers2016,Geers2017} and allows the direct use of crystal plasticity models developed for FEM codes (in particular Abaqus user material subroutine, \emph{umat}). To improve the accuracy of the microfields within the microstructure, a new projection operator is proposed based on the Fourier transform of discrete differentiation in the real space \cite{Willot2015}.

The robustness, accuracy and efficiency of the FFT framework proposed for micromechanics based fatigue simulation are assessed by comparing with finite element simulations. It is found that the macrosopic cyclic response of both standard FFT and FFT based on the discrete operator are almost identical to the FEM results. In the case of the local fields, small differences are found being this difference strongly reduced with the use of the discrete projection operator in the FFT problem. 

 The effect of the microfield differences between FFT and FEM in the fatigue life is analyzed by comparing the Fatigue Indicator Parameter (FIP) of the microstructure obtained from the extreme  values of the FIP distributions and averaging using different RVEs. It is shown that if the modified projection operator is used, the difference in the FIP value obtained using FFT respect FEM is below 7\% and therefore difference in fatigue estimation will be of the same order. Moreover, the introduction of hard particles in the polycrystalline RVEs is also feasible with accurate results thanks to the introduction of the discrete projection operator.These results confirms the validity of this modeling approach for predicting fatigue life. 

 Finally, the efficiency of this framework for simulating cyclic tests is examined and it is found that FFT is several times superior to FE (6 to 10 times faster for 128 voxel models) and allows to compute models with sizes not accessible using FEM. In summary, this framework combines the efficiency of FFT for solving large linear problems with an accuracy similar to FEM thanks to the discrete operator and a very efficient implicit integration of the non-linear behavior allowing to simulate complex RVEs and a large number of cycles, ideal conditions for micromechanics based fatigue life prediction.

\section*{Acknowledgments}
This investigation was supported by ITP (Industria de Turbo Propulsores. S. A.), the Comunidad de Madrid through the grant PEJD-2016/IND-2824 and the Spanish Ministry of Economy and Competitiveness through the project DPI2015-67667-C3-2-R. The authors thank Dr. A. Linaza and Dr. K. Ostolaza from ITP for their useful advises and to Dr. A.Cruzado (Texas A\&M), M. Spinola and Jun Lian Wang for their help in testing and evaluating the FFT code.

\appendix

\section{Coupling with Abaqus material user subroutines}\label{tangdev}
In this appendix, the adaptation of the Abaqus user material subroutine \emph{umat} to the FFT framework will be presented for a generic non-linear material with internal variables and under finite deformations. 

The FFT method needs to evaluate at each iteration the constitutive equation (eq. \ref{eq:constitutive}) for each point as function of the deformation gradient at current and previous time increments, $\mathbf{F}^{t_k}$ and $\mathbf{F}^{t_{k-1}}$ respectively, the last value of the internal variables, $\boldsymbol{\alpha}^{t_{k-1}}$, and the time step $\delta t$. These values are also inputs of a \emph{umat} subroutine and are passed from the FFT code to the \emph{umat} subroutine through the variables \verb|DFGRD1,DFGRD0,STATEV| and \verb|DTIME| respectively. After the evaluation, the FFT code needs to recover from the constitutive model the first Piola Kirchhoff stress $\mathbf{P}^{t_k}$, the value of the state variables at $t_k$ and the material tangent $\mathbb{K}$. The \emph{umat} subroutine provides the Cauchy stress at time $t_k$ ($\boldsymbol{\sigma}$) in the variable \verb|STRESS| so, the first Piola-Kirchhoff stress is computed from this value and $\mathbf{F}=\mathbf{F}^{t_k}$ using
\begin{equation}\label{fPK}
\mathbf{P}=J\boldsymbol{\sigma} \mathbf{F}^{-T}
\end{equation}
with $J=\det(\mathbf{F})$.

In the case of $\mathbb{K}$, an exact transformation of the consistent tangent matrix defined in the \emph{umat} (variable \verb|DDSDDE|) to the material tangent used by the FFT code is fundamental to preserve in FFT the convergence rate obtained with the user material in a finite element simulation. The material Jacobian that should be defined in a user subroutine at finite strain is the tangent modulus tensor for the Jaumann rate of the Kirchhoff stress, $\mathbb{C}^{ab}$, a fourth order tensor defined as
\begin{equation}\label{ABAtang}
\overset{\nabla}{\boldsymbol{\tau}}=
\dot{\boldsymbol{\tau}} - \mathbf{w} \cdot \boldsymbol{\tau}-\boldsymbol{\tau}\cdot\mathbf{w}^T=
J \ \mathbb{C}^{ab} : \mathbf{d}
\end{equation}
where $\boldsymbol{\tau}$ is the Kirchhoff stress, $\overset{\nabla}{}$ corresponds to the Jaumann rate and $\mathbf{w}$ and $\mathbf{d}$ are the spin tensor and stretch tensor respectively.  In the FFT framework, the material tangent needed to define the linear operator of the equilibrium in each iteration (eq. \ref{newlinop}) is defined as 
\begin{equation}\label{FFTtang}
\dot{\mathbf{P}} = \mathbb{K}:\dot{\mathbf{F}}
\end{equation}
and the objective then is to derive an explicit expression relating both tensors, $\mathbb{C}^{ab}$ and $\mathbb{K}$. 



Combining the definition of the material tangent in FFT, eq. (\ref{FFTtang}), and eq. (\ref{fPK}) it is obtained
\begin{equation}\label{proddivision}
\mathbb{K}=\frac{\partial \mathbf{P}}{\partial \mathbf{F}}=
\frac{\partial \left( J\boldsymbol{\sigma}\cdot\mathbf{F}^{-T}\right)}{\partial \mathbf{F}}=
\frac{\partial \left( \boldsymbol{\tau}\cdot\mathbf{F}^{-T}\right)}{\partial \mathbf{F}}.
\end{equation}
Expressing the previous expression (eq. \ref{proddivision}) in index notation and expanding the derivative of the product in two terms it is obtained
\begin{equation}\label{indexx}
K_{ijkl}=\frac{\partial P_{ij}}{\partial F_{kl}}=
\frac{\partial \left( \tau_{ip} F^{-1}_{jp}\right)}{\partial F_{kl}}=
\frac{\partial \tau_{ip}}{\partial F_{kl}} F^{-1}_{jp}+
\tau_{ip}\frac{\partial F^{-1}_{jp}}{\partial F_{kl}}.
\end{equation}

The second term of equation (\ref{indexx}), the derivative of the inverse of a tensor with respect to itself, is given by \ref{derivativeinv}.
\begin{equation}\label{derivativeinv}
\dfrac{\partial F^{-1}_{jp}}{\partial F_{kl}}=-
F^{-1}_{lp}F^{-1}_{jk}
\end{equation}

To obtain first term of equation (\ref{indexx}), the expression in derivatives of equation (\ref{ABAtang}) is multiplied by a small time increment  yielding into \ref{linearization}.
\begin{equation}\label{linearization}
\delta\boldsymbol{\tau} - \delta\mathbf{w} \cdot \boldsymbol{\tau}-\boldsymbol{\tau}\cdot\delta\mathbf{w}^T=
J \ \mathbb{C}^{ab} : \delta\mathbf{d}
\end{equation}
and reordering the equation, the Abaqus tangent can be written as
\begin{equation}\label{dtau}
\delta\boldsymbol{\tau} =
J \ \mathbb{C}^{ab} : \delta\mathbf{d} +\delta\mathbf{w} \cdot \boldsymbol{\tau} +\boldsymbol{\tau}\cdot\delta\mathbf{w}^T
\end{equation}
where $\delta\mathbf{d}$ and $\delta\mathbf{w}$ are obtained as function of $\mathbf{F}$ and $\delta\mathbf{F}$ by
\begin{equation}\label{basic}
\delta \mathbf{d}=
\frac{1}{2}\left[\delta\mathbf{F}\cdot\mathbf{F}^{-1}+\left(\delta\mathbf{F}\cdot\mathbf{F}^{-1}\right)^{T}\right]
\end{equation}
and
\begin{equation}\label{basic2}
\delta \mathbf{w}=
\frac{1}{2}\left[\delta\mathbf{F}\cdot\mathbf{F}^{-1}-\left(\delta\mathbf{F}\cdot\mathbf{F}^{-1}\right)^{T}\right].
\end{equation}
Replacing these expressions eqs. (\ref{basic}) and (\ref{basic2}) into equation (\ref{dtau}), it is obtained
\begin{equation}\label{dtauin}
\begin{aligned}
\delta\boldsymbol{\tau} = J \ \mathbb{C}^{ab} : \frac{1}{2}\left[\delta\mathbf{F}\cdot\mathbf{F}^{-1} + \left(\delta\mathbf{F}\cdot\mathbf{F}^{-1}\right)^{T}\right] + 
\\ 
\frac{1}{2}\left[\delta\mathbf{F}\cdot\mathbf{F}^{-1}-\left(\delta\mathbf{F}\cdot\mathbf{F}^{-1}\right)^{T}\right] \cdot \boldsymbol{\tau} +
\\
\boldsymbol{\tau}\cdot\frac{1}{2}\left[\delta\mathbf{F}\cdot\mathbf{F}^{-1}-\left(\delta\mathbf{F}\cdot\mathbf{F}^{-1}\right)^{T}\right]^T.
\end{aligned}
\end{equation}

%

Now, the Kirchhoff stress is also linearized  respect the perturbation of the deformation gradient
\begin{equation}\label{lintau}
\delta\boldsymbol{\tau} =
\frac{\delta\boldsymbol{\tau}}{\delta\mathbf{F}} : \delta\mathbf{F}
\text{ in index }
\delta \tau_{ip} =
\frac{\partial \tau_{ip}}{\partial F_{kl}} \delta F_{kl}.
\end{equation}
If equation \ref{dtauin} is written in index notation, the resulting expression reads as
\begin{equation}\label{indexini}
\begin{aligned}
\delta\tau_{ip} = \frac{J}{2} C^{ab}_{ipkm} \delta F _{kl} F^{-1}_{lm} + \frac{J}{2} C^{ab}_{ipmk}  F^{-1}_{lm} \delta F _{kl} + 
\\ 
\frac{1}{2} I_{ipkq} \delta F_{kl} F^{-1}_{lm} \tau_{mq} -
\frac{1}{2} I_{ipmq} F^{-1}_{lm} \delta F_{kl} \tau_{kq} +
\\
\frac{1}{2} I_{ipqk} \tau_{qm} F^{-1}_{lm} \delta F_{kl} -
\frac{1}{2} I_{ipqm} \tau_{qk} F^{-1}_{lm} \delta F_{kl}
\end{aligned}
\end{equation}
and comparing the terms of the last two equations (\ref{lintau}) and (\ref{indexini}), considering the minor symmetries of $\mathbb{C}^{ab}$ and the symmetry of $\boldsymbol{\tau}$, the resulting expression is:

\begin{equation}
\begin{aligned}
\frac{\partial \tau_{ip}}{\partial F_{kl}} =
J \ C^{ab}_{ipkm} F^{-1}_{lm} +
\\
\frac{1}{2} \delta_{ik} \delta_{pq} F^{-1}_{lm} \tau_{mq} -
\frac{1}{2} \delta_{im} \delta_{pq} F^{-1}_{lm} \tau_{kq} +
\\
\frac{1}{2} \delta_{iq} \delta_{pk} F^{-1}_{lm} \tau_{qm} -
\frac{1}{2} \delta_{iq} \delta_{pm} F^{-1}_{lm} \tau_{qk},
\end{aligned}
\end{equation}
which can be simplified to
\begin{equation}
\label{tau_F_function_Cab}
\begin{aligned}
\frac{\partial \tau_{ip}}{\partial F_{kl}} =
J \ C^{ab}_{ipkm} F^{-1}_{lm} +
\\
\frac{1}{2} \delta_{ik} F^{-1}_{lm} \tau_{mp} -
\frac{1}{2} F^{-1}_{li} \tau_{kp} +
\\
\frac{1}{2} \delta_{pk} F^{-1}_{lm} \tau_{im} -
\frac{1}{2} F^{-1}_{lp} \tau_{ik}.
\end{aligned}
\end{equation}

Finally, introducing eq. (\ref{tau_F_function_Cab}) in the definition of the material tangent (eq. \ref{indexx}) the final expression for the material tangent used in the FFT approach as function of the Abaqus tangent is given by

\begin{equation}
\begin{aligned}
K_{ijkl}= &
( J \ C^{ab}_{ipkm} F^{-1}_{lm} +
\frac{1}{2} \delta_{ik} F^{-1}_{lm} \tau_{mp} - \\
& \frac{1}{2} F^{-1}_{li} \tau_{kp} +
\frac{1}{2} \delta_{pk} F^{-1}_{lm} \tau_{im}-\frac{1}{2} F^{-1}_{lp} \tau_{ik})F^{-1}_{jp}-
\tau_{ip} F^{-1}_{lp} F^{-1}_{jk}.
\end{aligned}
\end{equation}

\bibliographystyle{unsrt}
\bibliography{}
\end{document}